\documentclass[%
 reprint,
superscriptaddress,
amsmath,amssymb,aps,prl,floatfix,multicol,nobibnotes,showkeys]{revtex4-2}

\usepackage{bbm, bm, mathtools, hyperref, graphicx, xcolor, float} 

\begin{document}

\title[IST] {Fractional Integrable and Related Discrete Nonlinear Schr\"odinger Equations}

\author{Mark J. Ablowitz}
\address{Department of Applied Mathematics, University of Colorado, Boulder, Colorado 80309, U.S.A.}
\thanks{}

\author{Joel B. Been\footnote[10]{here}}
\email[Corresponding author: ]{joelbeen@mines.edu}
\address{Department of Applied Mathematics and Statistics, Colorado School of Mines, Golden, Colorado 80401, U.S.A.}
\address{Department of Physics, Colorado School of Mines, Golden, Colorado 80401, U.S.A.}

\author{Lincoln D. Carr}
\address{Department of Applied Mathematics and Statistics, Colorado School of Mines, Golden, Colorado 80401, U.S.A.}
\address{Department of Physics, Colorado School of Mines, Golden, Colorado 80401, U.S.A.}
\address{Quantum Engineering Program, Colorado School of Mines, Golden, Colorado 80401, U.S.A.}
\thanks{}

\begin{abstract}
    
    Integrable fractional equations such as the fractional Korteweg-deVries and nonlinear Schr\"odinger equations are key to the intersection of nonlinear dynamics and fractional calculus. In this manuscript, the first discrete/differential difference equation of this type is found, the fractional integrable discrete nonlinear Schr\"odinger equation. This equation is linearized; special soliton solutions are found whose peak velocities exhibit more complicated behavior than other previously obtained fractional integrable equations. This equation is compared with the closely related fractional \emph{averaged} discrete nonlinear Schr\"odinger equation which has simpler structure than the integrable case. For positive fractional parameter and small amplitude waves, the soliton solutions of the integrable and averaged equations have similar behavior. 



\end{abstract}


\maketitle

\section{Introduction}

Integrable systems play a central role in nonlinear dynamics because they provide exactly solvable models for important physical systems. 
Notable examples of integrable equations are the Korteweg-deVries (KdV), applicable to shallow water waves, plasma physics, and lattice dynamics among others \cite{KDV,Ablowitz2,Ablowitz2011}, and the nonlinear Schr\"odinger (NLS) equation, which finds applications in nonlinear optics, Bose-Einstein condensates, spin waves in ferromagnetic films, plasma physics, water waves, etc. \cite{Ablowitz2,Ablowitz2011,bronski_2001,boardman_1994}. These integrable nonlinear evolution equations have an infinite number of conservation laws and soliton solutions \cite{Ablowitz1}. Solitons, the fundamental solutions of such equations, are stable, localized nonlinear waves which propagate without dispersing and interact elastically with other solitons. Nonlinear integrable evolution equations have these surprising properties because of their deep mathematical structure described by the inverse scattering transform (IST).

IST is a method of solving nonlinear equations which generalizes Fourier transforms. It solves these equations in three steps: mapping the initial condition into scattering space, evolving the intial data in scattering space in time, and mapping the evolved scattering data back to physical space; i.e., inverse scattering. This process gives the solution to nonlinear equations solvable by IST in terms of linear integral equations; such nonlinear equations are called integrable. Recently, we used the mathematical structure of IST associated with for the Korteweg-deVries (KdV) and nonlinear Schr\"odinger (NLS) equations to develop a method of finding and solving the fractional KdV (fKdV) and fractional NLS (fNLS) equations \cite{fKdV_fNLS}. We also showed that this method could be applied to find fractional extensions of the modified KdV, sine-Gordon, and sinh-Gordon equations \cite{fmKdV}. These equations represent the first known fractional integrable nonlinear evolution equations with smooth (physical) solutions and deeply connect the fields of nonlinear dynamics and fractional calculus. 

Fractional calculus is a mathematical structure originally designed to define non-integer derivatives and integrals. It has sense become an effective way of modeling many physical processes that exist in multi-scale media \cite{west_colloquium,zhong2016spatiotemporal} or exhibit non-Gaussian statistics or power law behavior \cite{random_walk,what_is_frac_lap,frac_stoch}. A particularly important example is anomalous diffusion, where the mean squared displacement is proportional to $t^{\alpha}$, $\alpha>0$ \cite{west_1987,random_walk,west_1997,anomalous_issue}. Transport that follows this rule has been observed in biology \cite{saxton_2007,bronstein_2009,weigel_2011,regner_2013}, amorphous materials \cite{scher_1975,pfister_1977,gu_1996}, porous media \cite{benson_2000,benson_2001,meerschaert_2008,frac_porous}, climate science \cite{koscielny_1998}, and attenuation in materials \cite{power_law_attn} amongst others. As we have shown, the merger of fractional and nonlinear characteristics in integrable equations such as fKdV and fNLS predict \emph{anomalous dispersion}, where the velocity and amplitude of solitonic solutions are related by a power law \cite{fKdV_fNLS}.

In this article, we demonstrate how the method introduced in Ref. \cite{fKdV_fNLS} can be applied to discrete (or differential-difference) systems to define integrable discrete fractional nonlinear evolution equations by presenting a fractional generalization of the integrable discrete nonlinear Schr\"odinger (IDNLS) equation. We do this by demonstrating the three key mathematical ingredients of our method --- IST, power law dispersion relations, and completeness relations --- for the Ablowitz-Ladik (AL) discrete scattering problem. In the linear limit, the fractional IDNLS (fIDNLS) equation is a discretization of the fractional Schr\"odinger equation which was derived with Feynman path integrals over L\'evy flights \cite{laskin2000fractional,laskin2018}.

The KdV equation was the first equation shown to be solvable by IST in Ref. \cite{KDV1967}; it was soon followed by the NLS equation in Ref. \cite{ZS72}. 
These two equations were then found to be contained in a general class of equations solvable by IST when associated to the Ablowitz-Kaup-Newell-Segur (AKNS) system \cite{AKNS,Ablowitz2011}. Shortly thereafter IST was used to solve families of discrete (or differential difference) problems like the self-dual network \cite{ablowitz1975nonlinear}. In particular, it was discovered that the AKNS system could be discretized while maintaining integrability, leading to the AL scattering problem which was used to solve a family of discrete nonlinear evolution equations \cite{ablowitz1976nonlinear}. This family contained important discrete evolution equations --- continuous in time but discretized in space --- such as integrable discretizations of the nonlinear Schr\"odinger, KdV, modified KdV, and sine Gordon equations. Further, this family of equations was shown to have soliton solutions and an infinite number of conservation laws \cite{ablowitz1976nonlinear}.

We derive the fIDNLS equation from the AL scattering problem using three key components: linear dispersion relations, completeness relations, and IST. IST is used to linearized the equation and obtain special soliton solutions.

We also show how the characteristics of the fractional IDNLS (fIDNLS) equation reach beyond integrability by comparing the one-soliton solution of the fIDNLS equation to the solitary wave solution of the fractional averaged discrete nonlinear Schr\"odinger (fADNLS) equation. This equation is a different fractional generalization of the IDNLS equation in which the linear second order difference is replaced by the discrete fractional Laplacian \cite{molina2020two,ciaurri2015fractional,huang2014numerical,ciaurri2017harmonic}. The fADNLS equation can be understood as a discretization of a fractional NLS equation involving the Riesz derivative which has been extensively studied in, e.g., \cite{iomin2021fractional,malomed2021optical,qiu2020stabilization,li2021symmetry,al2018high}; it is also is also closely related to the (likely) non-integrable fractional DNLS equation, recently studied in \cite{molina2020two,PhysRevE.103.042214}. Though the fADNLS equation is likely not integrable to our knowledge (apart from the limiting case when fADNLS reduces to IDNLS), the similarity between the two equations suggests that some of the physical predictions of fractional integrable equations are shared by equations which are simpler to realize computationally.

\section{The discrete fractional linear Schr\"odinger equation}
Consider the family of discrete linear evolution equations
\begin{align}
    \partial_{t} q_{n} + \gamma(-\Delta_{n}) q_{n} = 0 \label{eqn:discrete_linear_evolution_equation_family}
\end{align}
for the function $q_{n}(t)$ which depends on the discrete variable $n\in\mathbb{Z}$ and the continuous variable $t\in\mathbb{R}$. Here, $\gamma$ is a sufficiently regular function of the discrete laplacian, $-\Delta_{n}$, defined by 
\begin{align}
    (-\Delta_{n}) q_{n}(t) = \frac{1}{h^2}\big(- q_{n+1}(t) + 2 q_{n}(t) - q_{n-1}(t)\big) \label{eqn:discrete_laplacian_definition}
\end{align}
where $h$ is the distance between lattice sites. Using the Z-transform, which is equivalent to the discrete Fourier transform, the solution to Eq. (\ref{eqn:discrete_linear_evolution_equation_family}) can be explicitly written as
\begin{align}
    q_{n}(t) = \frac{1}{2\pi} \int_{-\pi/h}^{\pi/h} dk \hat{q}(k,0) e^{i k n h - \gamma(4 \sin^2(kh/2)/h^2)t} \label{eqn:discrete_linear_evolution_equation_solution}
\end{align}
where $\hat{q}(k,0) = h \sum_{n = -\infty}^{\infty} q_{n}(0) e^{-i k n h}$ is the Z-transform of $q_{n}(t)$ at $t = 0$ and $4 \sin^2(kh/2)/h^2$ is the Fourier symbol of $-\Delta_{n}$. Note that the Z-transform is often written in terms of $z$ with the substitution $z = e^{-i k h}$ where integration in $k$ becomes integration with respect to $z$ on the unit circle. If we choose $\gamma$ to be power law, then Eq. (\ref{eqn:discrete_linear_evolution_equation_family}) becomes a fractional discrete equation in terms of the discrete fractional laplacian. For example, if we put $\gamma(-\Delta_{n}) = - i (-\Delta_{n})^{1+\epsilon}$, $|\epsilon|<1$, then we obtain the linear fractional discrete Schr\"odinger equation
\begin{align}
    i \partial_{t} q_{n} + (-\Delta_{n})^{1+\epsilon} q_{n} = 0. \label{eqn:fractional_discrete_linear_schrodinger_equation}
\end{align}
Here, $(-\Delta_{n})^{1+\epsilon}$ is the discrete fractional laplacian of order $1+\epsilon$ which is defined in terms of its Fourier symbol $[4 \sin^2(hk/2)/h^2]^{1+\epsilon}$ and the Z-transform/discrete Fourier transform as
\begin{align}
    (-\Delta_{n})^{1+\epsilon} q_{n} = \frac{1}{2\pi}\!\! \int_{-\pi/h}^{\pi/h}\!\! dk \hat{q}(k)e^{i k n h} [4 \sin^2(k h/2)/h^2]^{1+\epsilon}. \label{eqn:discrete_fractional_laplacian_definition}
\end{align}
Notice that the $k$ integral above can be evaluated to express the discrete fractional laplacian as a summation over $m$ of $q_{m}$ multiplied by a weight vector. The solution to Eq. (\ref{eqn:fractional_discrete_linear_schrodinger_equation}) can still be written in the form Eq. (\ref{eqn:discrete_linear_evolution_equation_solution}) with
\begin{align}
    \gamma(4 \sin^2(k h/2)/h^2) = - i [4 \sin^2(k h/2)/h^2]^{1+\epsilon} \notag
\end{align} and, because $4 \sin^2(k h/2)/h^2$ is real and positive, the solution to equation (\ref{eqn:discrete_linear_evolution_equation_family}) with this choice of $\gamma$ is well posed. In defining and solving the linear fractional discrete Schr\"odinger equation, we used a power law dispersion relation, ingredient $1$ of our method, and we defined the fractional operator using completeness of the discrete Fourier transform/Z-transform, ingredient $2$. Then we solve the equation by the inverse discrete Fourier transform, the analog of ingredient $3$.

\section{The fractional integrable discrete Schr\"odinger equation}
To develop the fIDNLS equation, the integrable nonlinear analog of Eq. (\ref{eqn:fractional_discrete_linear_schrodinger_equation}), and solve it, we apply the three key ingredients of our method, starting with writing the equation in terms of a linear dispersion relation. Note that $h = 1$ is taken in this section without loss of generality; to recover the scaling factor for $h \ne 1$, replace $q_{n}$ by $h q_{n}$ and $r_{n}$ by $h r_{n}$.

As in the linear case, Eq. (\ref{eqn:discrete_linear_evolution_equation_family}), we have a family of nonlinear evolution equations for the solutions $q_{n}(t)$ and $r_{n}(t)$ \cite{gerdjikov1984expansions}, see also \cite{ChiuLadik1977},
\begin{align}
    \sigma_{3} \frac{d\mathbf{u}_{n}}{dt} + \gamma(\Lambda_{+}) \mathbf{u}_{n} = 0, ~~ \mathbf{u}_{n} = \left(q_{n}, -r_{n} \right)^{T} \label{eqn:AL_general_evolution_equation}
\end{align}
where $T$ represents transpose, $\sigma_{3} = \text{diag}(1,-1)$, and $\Lambda_{+}$ is
\begin{align}
    &\Lambda_{+} \mathbf{x}_{n} = h_{n} \begin{pmatrix}
    E_{n}^{+} & 0 \\ 0 & E_{n}^{-}
    \end{pmatrix} \begin{pmatrix} x_{k}^{(1)} \\ x_{k}^{(2)} \end{pmatrix} \\
    &+ \begin{pmatrix}
    q_{n} \sum_{n-1}^{+} r_{k-1} & q_{n} \sum_{n-2}^{+} q_{k+1} \\
    - r_{n} \sum_{n-1}^{+} r_{k-1} & - r_{n} \sum_{n-2}^{+} q_{k+1}
    \end{pmatrix} \begin{pmatrix} x_{k}^{(1)} \\ x_{k}^{(2)} \end{pmatrix} \\
    &+ h_{n} \begin{pmatrix}
    q_{n+1} \sum_{n+1}^{+} \frac{r_{k}}{h_{k}} & q_{n+1} \sum_{n+1}^{+} \frac{q_{k}}{h_{k}} \\
    -r_{n-1} \sum_{n}^{+} \frac{r_{k}}{h_{k}} & -r_{n-1} \sum_{n}^{+} \frac{q_{k}}{h_{k}}
    \end{pmatrix} \begin{pmatrix} x_{k}^{(1)} \\ x_{k}^{(2)} \end{pmatrix}
\end{align}
where $h_{n} = 1 - r_{n} q_{n}$, $\sum_{n}^{+} = \sum_{k = n}^{\infty}$, and $E_{n}^{\pm} x^{(q)}_{k} = x^{(q)}_{n\pm1}$ with $q = 1,2$. The inverse of this operator is
\begin{align}
    &\Lambda_{+}^{-1} \mathbf{x}_{n} = h_{n}\begin{pmatrix}
     E_{n}^{-} & 0 \\
    0 & E_{n}^{+}
    \end{pmatrix}\begin{pmatrix} x_{k}^{(1)} \\ x_{k}^{(2)} \end{pmatrix} \\
    &+ \begin{pmatrix}
    - q_{n} \sum_{n}^{+} r_{k+1} & - q_{n} \sum_{n+1}^{+} q_{k-1} \\
    r_{n} \sum_{n-1}^{+} r_{k+1} & r_{n} \sum_{n}^{+} q_{k-1}
    \end{pmatrix}\begin{pmatrix} x_{k}^{(1)} \\ x_{k}^{(2)} \end{pmatrix} \\
    &+ h_{n} \begin{pmatrix}
    -q_{n-1} \sum_{n}^{+} \frac{r_{k}}{h_{k}} & - q_{n-1} \sum_{n}^{+} \frac{q_{k}}{h_{k}} \\
    r_{n+1} \sum_{n+1}^{+} \frac{r_{k}}{h_{k}} & -r_{n+1} \sum_{n+1}^{+} \frac{q_{k}}{h_{k}}
    \end{pmatrix}\begin{pmatrix} x_{k}^{(1)} \\ x_{k}^{(2)} \end{pmatrix}.
\end{align}
Here, $\gamma$ is a sufficiently regular function of the operator $\Lambda_{+}$ and is connected with the linearized dispersion relation. Specifying this dispersion relation, or $\gamma$ directly, picks out particular equations from this family. For example, if we take
\begin{align}
    \gamma(\Lambda_{+}) = - i (2 - \Lambda_{+} - \Lambda_{+}^{-1}) \notag
\end{align}
and let $r_{n} = \mp q_{n}^{*}$, then we obtain the IDNLS equation
\begin{align}
    i \partial_{t} q_{n} + \Delta_{n} q_{n} \pm |q_{n}|^2 (q_{n+1} + q_{n-1}) = 0. \label{eqn:integrable_discrete_NLS}
\end{align}
We can relate $\gamma$ to the dispersion relation of the linearization of (\ref{eqn:AL_general_evolution_equation}) by considering the linear limit $q_n\to0$. In this limit, we have
\begin{align}
    \Lambda_{+} \to \begin{pmatrix} E_{n}^{+} & 0 \\ 0 & E_{n}^{-} \end{pmatrix} \equiv \mathbf{D}_{n},
\end{align}
so the linearization of the nonlinear evolution equation is
\begin{align}
    \sigma_{3}\frac{d \mathbf{u_{n}}}{dt} + \gamma(\mathbf{D}_{n}) \mathbf{u}_{n} = 0. \label{eqn:linear_AL_general_evolution_equation}
\end{align}
Because $\mathbf{D}_{n}$ is a diagonal matrix, we have
\begin{align}
    \gamma(\mathbf{D}_{n}) = \begin{pmatrix} \gamma(E_{n}^{+}) & 0 \\ 0 & \gamma(E_{n}^{-}) \end{pmatrix}.
\end{align}
Taking the first component of (\ref{eqn:linear_AL_general_evolution_equation}) with
\begin{align}
    q_{n} = z^{2 n} e^{ - i \omega(z) t} \notag
\end{align} gives
\begin{align}
    \gamma(z^2) = i \omega(z).
\end{align}
Therefore, by specifying the linear limit of the nonlinear evolution equation, we obtain the nonlinear equation itself. To define the fIDNLS equation, we choose the linear limit to be the discrete linear fractional Schr\"odinger equation in (\ref{eqn:fractional_discrete_linear_schrodinger_equation}), which gives the dispersion relation $\omega(z) = -(2 - z^{2} - z^{-2})^{1+\epsilon}$ and, hence, $\gamma(z^2) = - i (2 - z^{2} - z^{-2})^{1+\epsilon}$. So, the fIDNLS equation is
\begin{align}
    i \partial_{t} \mathbf{u}_{n} + (2 - \Lambda_{+} - \Lambda_{+}^{-1})^{1+\epsilon} \mathbf{u}_{n}(t) = 0. \label{eqn:fIDNLS}
\end{align}
In fact, by choosing $\gamma(z^2) = - i (2 - z^{2} - z^{-2})^{m+\epsilon}$, for integer $m$, we generate a hierarchy of fractional equations
\begin{align}
    i \partial_{t} \mathbf{u}_{n} + (2 - \Lambda_{+} - \Lambda_{+}^{-1})^{m+\epsilon} \mathbf{u}_{n}(t) = 0. \label{eqn:fIDNLS_hierarchy}
\end{align}
It can be shown that the limit of (\ref{eqn:fIDNLS}) as $\epsilon\to0$ is the IDNLS equation (\ref{eqn:integrable_discrete_NLS}). Notice that to define the fIDNLS equation, we used a power law dispersion relation, ingredient 1 of the method. However, this dispersion relation leads to the operator $(2 - \Lambda_{+} - \Lambda_{+}^{-1})^{1+\epsilon}$ the meaning of which is currently unclear. To define this operator, we will need to use the 2nd ingredient: appropriate completeness relations. The third ingredient will be making use of IST to find solutions of the fIDNLS equation.

\section{Completeness of Squared Eigenfunctions and Fractional Operators}
In this section we define the fIDNLS equation in (\ref{eqn:fIDNLS}) and, in fact, any equation of the form (\ref{eqn:discrete_linear_evolution_equation_family}) that is well-posed in physical space. We do this using the observation that $\gamma(\Lambda_{+})$ is a multiplication operator when acting on the eigenfunctions of $\Lambda_{+}$ and the fact that the eigenfunctions of $\Lambda_{+}$ are complete. This result is known as completeness of squared eigenfunctions, and is the second ingredient in our method. The resulting representation of $\gamma(\Lambda_{+})$ will be similar to that of the discrete fractional laplacian in (\ref{eqn:discrete_fractional_laplacian_definition}). The eigenfunctions of $\Lambda_{+}$ are $\mathbf{\Psi}_{n}(z)$ and $\overline{\mathbf{\Psi}}_{n}(z)$ each with eigenvalue $z^2$ (note that time $t$ is suppressed throughout this section). Therefore, the operation of $\gamma(\Lambda_{+})$ on these eigenfunctions is given by
\begin{align}
    \gamma(\Lambda_{+}) \mathbf{\Psi}_{n} = \gamma(z^2) \mathbf{\Psi}_{n}, ~~ \gamma(\Lambda_{+}) \overline{\mathbf{\Psi}}_{n} = \gamma(z^2) \overline{\mathbf{\Psi}}_{n}. \label{eqn:gamma_eigenfunctions}
\end{align}
Because $\Lambda_{+}$ is not a self-adjoint operator, completeness of squared eigenfunctions involves both $\mathbf{\Psi}_{n}$, $\overline{\mathbf{\Psi}}_{n}$ and the adjoint functions $\mathbf{\Psi}^{A}_{n}$, $\overline{\mathbf{\Psi}}^{A}_{n}$ where
\begin{align}
    \gamma(\Lambda_{+}^{A}) \mathbf{\Psi}_{n}^{A} = \gamma(z^2) \mathbf{\Psi}_{n}^{A}, ~~ \gamma(\Lambda_{+}^{A}) \overline{\mathbf{\Psi}}^{A}_{n} = \gamma(z^2) \overline{\mathbf{\Psi}}^{A}_{n} \label{eqn:gamma_adjoint_eigenfunctions}
\end{align}
and $\Lambda_{+}^{A}$ is the adjoint, with respect to $\ell^{2}(\mathbb{Z})\times\ell^{2}(\mathbb{Z})$, of $\Lambda_{+}$. The eigenfunctions and adjoint eigenfunctions can be written in terms of solutions to the Ablowitz-Ladik scattering problem which is a $2\times2$ eigenvalue problem fo the discrete vector-valued function $\mathbf{v}_{n} = (v_{n}^{(1)},v_{n}^{(2)})^{T}$
\begin{align}
    \mathbf{v}_{n+1} = \begin{pmatrix} z & q_{n} \\ r_{n} & z^{-1} \end{pmatrix} \label{eqn:AL_scattering_problem}
\end{align}
where $q_n$ and $r_n$ act as potentials and $z$ is an eigenvalue. Through this association, one can solve the family of nonlinear evolution equations in (\ref{eqn:AL_scattering_problem}) (see Appendix for more details).

In \cite{gerdjikov1984expansions}, it was shown that the arbitrary discrete function $\mathbf{H}_{n} = \left(H_{n}^{(1)}, H_{n}^{(2)}\right)^{T}\in l_{1}$ can be written as
\begin{align}
    \mathbf{H}_{n} = \sum_{p = 1}^{2} \oint_{S^{(p)}} \frac{dz}{z} f^{(p)}(z) \sum_{m = -\infty}^{\infty} \mathbf{G}_{n,m}^{(p)}(z)\, \mathbf{H}_{m} \label{eqn:AL_completness_relation}
\end{align}
where $S^{(1)} = S_{R}$ ($S^{(2)} = S_{\delta}$) is a circular contour evaluated counterclockwise centered at the origin of radius $R>1$ ($\delta<1$) such that all of the singularities of the integrand are inside (outside) of the contour and
\begin{align}
    \mathbf{G}_{n,m}^{(1)}(z) &= \mathbf{\Psi}_{n}(z) \mathbf{\Psi}^{A}_{m}(z)^{T}/h_{n}, ~ f^{(1)}(z) = \frac{i}{2 \pi a^2(z)} \\
    \mathbf{G}_{n,m}^{(2)}(z) &= \overline{\mathbf{\Psi}}_{n}(z) \overline{\mathbf{\Psi}}^{A}_{m}(z)^{T}/h_{n}, ~ f^{(2)}(z) = \frac{-i}{2 \pi \overline{a}^2(z)}
\end{align}
with $h_{n} = 1 - r_{n} q_{n}$. The eigenfunctions $\mathbf{\Psi}_{n}(z)$,$\mathbf{\Psi}^{A}_{n}(z)$,$\overline{\mathbf{\Psi}}_{n}(z)$,$\overline{\mathbf{\Psi}}^{A}_{n}(z)$ (see appendix) and scattering data $a(z)$, $\overline{a}(z)$ are defined in terms of solutions to the Ablowitz-Ladik scattering problem (see Appendix). With this completeness relation, and the operation of $\gamma(\Lambda_{+})$ on $\mathbf{\Psi}_{n}$ and $\overline{\mathbf{\Psi}}_{n}$ in Eq. (\ref{eqn:gamma_eigenfunctions}), we have
\begin{align}
    \gamma(\Lambda_{+}) \mathbf{H}_{n} = \sum_{p = 1}^{2} \oint_{S^{(p)}} \! \frac{dz}{z} f^{(p)}(z) \gamma(z^2) \! \sum_{m = -\infty}^{\infty}\! \mathbf{G}_{n,m}^{(p)}(z) \mathbf{H}_{m}. \label{eqn:Lambda_spectral_representation}
\end{align}
Therefore, the nonlinear evolution equation in (\ref{eqn:AL_general_evolution_equation}) can be explicitly characterized in physical space as
\begin{align}
    \sigma_{3} \frac{d\mathbf{u}_{n}}{dt} = - \sum_{p = 1}^{2} \oint_{S^{(p)}} \! \frac{dz}{z} f^{(p)}(z) \gamma(z^2) \! \sum_{m = -\infty}^{\infty} \! \mathbf{G}_{n,m}^{(p)}(z) \mathbf{u}_{m}. \label{eqn:AL_spectral_general_evolution_equation}
\end{align}
In particular, if we put $\gamma(z^2) = - i (2 - z^{2} - z^{-2})^{1+\epsilon}$ and $r_{n} = \mp q_{n}^{*}$, the fIDNLS equation is the first component of (\ref{eqn:AL_spectral_general_evolution_equation}). Using the symmetries of the eigenfunctions (see appendix), this is
\begin{align}
    i \partial_t q_{n} = \sum_{p = 1}^{2} \oint_{S^{(p)}} \! \frac{dz}{z} f^{(p)}(z) \gamma(z^2) \! \sum_{m = -\infty}^{\infty} \! g_{n,m}^{(p)}(z)
\end{align}
with
\begin{align}
    &g_{n,m}^{(1)}(z) = - i\frac{\nu_{n} \nu_{m}}{h_{n}} \psi^{(1)}_{n}(z) \psi^{(1)}_{n+1}(z) \\
    &\cdot\left( \phi^{(2)}_{m}(z) \phi^{(2)}_{m+1}(z) q_{m} \mp \phi^{(1)}_{m}(z) \phi^{(1)}_{m+1}(z) q_{m}^{*} \right) \notag \\
    &g_{n,m}^{(2)}(z) = - i\frac{\nu_{n} \nu_{m}}{h_{n}} \left(\psi^{(2)}_{n}(1/z^{*}) \psi^{(2)}_{n+1}(1/z^{*}) \right)^{*} \\
    &\cdot\left( \phi^{(1)}_{m}(1/z^{*}) \phi^{(1)}_{m+1}(1/z^{*}) q_{m}^{*} \mp \phi^{(2)}_{m}(1/z^{*}) \phi^{(2)}_{m+1}(1/z^{*})1 q_{m} \right)^{*} \notag
\end{align}
where $t$ has been suppressed.

In the appendix we show how this equation can be linearized via Gel'fand-Levitan-Marchenko type summation equations. After long time the kernel of the summation equation contains only discrete spectra, i.e., the soliton solutions. Multisoliton solutions can be found by standard methods.

\section{Solitons and Solitary wave solutions of the  fIDNLS and fADNLS equations}
The fIDNLS equation in (\ref{eqn:fIDNLS}) is not the only fractional generalization of the IDNLS equation in (\ref{eqn:integrable_discrete_NLS}). A simpler generalization is to replace the discrete laplacian $-\Delta_{n}$ in (\ref{eqn:integrable_discrete_NLS}) with the discrete fractional laplacian $(-\Delta_{n})^{1+\epsilon}$ defined in (\ref{eqn:discrete_fractional_laplacian_definition}) to give the fractional averaged DNLS (fADNLS) equation
\begin{align}
    i \partial_{t} q_{n} + (-\Delta_{n})^{1+\epsilon} q_{n} \pm |q_{n}|^2 (q_{n+1} + q_{n-1}) = 0. \label{eqn:fADNLS}
\end{align}
Notice that in the figure captions we refer to the fIDNLS equation as the fractional integrable equation and the fADNLS equation as the fractional averaged equation.

The fADNLS equation is not known to be integrable, but in the limit $\epsilon\to0$, it becomes the IDNLS equation, Eq. (\ref{eqn:integrable_discrete_NLS}), which is integrable; therefore, we expect Eq. (\ref{eqn:fADNLS}) to have some similarity the fIDNLS equation. To characterize this similarity, we will compare the solitons and solitary waves predicted by these equations. The fIDNLS equation has an exact one-soliton solution, derivable by the IST. 
To find the solitary wave solutions to the fADNLS equation we use the same initial condition as that of the fIDNLS equation. 

Even though this 
solitary wave initially deforms from the exact secant profile, emitting radiation in the process, its solutions have nearly constant velocity, propagate with nearly constant amplitude, and have comparable velocities to the fIDNLS equation in certain regimes. These integrable-like properties of this equation are stronger for positive $\epsilon$ than negative $\epsilon$ and stronger for smaller wave amplitudes than larger wave amplitudes. Soliton solutions to the fIDNLS equation can be derived using the IST (see appendix and \cite{Ablowitz3}); they are of the form
\begin{align}
    q_{n}(t) &= \frac{\sinh{(2 \eta h)}}{h} e^{2 i \left(v_{i}(z_{1}^2) t - \xi h n\right) - i \left(\psi - \pi/2\right)} \label{eqn:fIDNLS_one_soliton} \\
    &\times\text{sech}\left( 2 \eta h (n - n_{0}) - 2 v_{r}(z_{1}^2) t \right) \notag
\end{align}
where $v_{i}(z_{1}^2) = \frac{1}{2} \text{Im}\gamma(z_{1}^2)$, $v_{r}(z_{1}^2) = \frac{1}{2} \text{Re}\gamma(z_{1}^2)$, and $z_{1} = e^{h(\eta - i\xi)}$. Here we choose $\gamma(z_{1}^2) = - i (2 - z_{1}^{2} - z_{1}^{-2})^{1+\epsilon}$ in accordance with Eq. (\ref{eqn:fIDNLS}) though (\ref{eqn:fIDNLS_one_soliton}) holds for all sufficiently regular $\gamma$. The free parameters in (\ref{eqn:fIDNLS_one_soliton}) are $\epsilon$, $h$, $\eta$, $\xi$, $\psi$, and $n_{0}$.

To find the localized wave solutions to the fADNLS equation, we numerically evolved the equation at discrete time steps $\{t_{m}\}_{m = 0}^{M}$ with $t_{0} = 0$ using a Fourier split-step scheme. The initial condition $q_{n}(t_{0}) = q_{n}(0)$ is given by (\ref{eqn:fIDNLS_one_soliton}) with $t = 0$. The Fourier split-step scheme propagates the approximation from $t_{m}$ to $t_{m+1}$ by separately evaluating the linear and nonlinear parts of the equation; cf. Refs. \cite{TAHA1984203,Hardin1973ApplicationOT,Sinkin:03}. Explicitly, we compute
\begin{align}
    q_{n}(t_{m+1}) = e^{\!- i \Delta t_{m} \mathcal{L}/2} e^{i \int_{t_{m}}^{t_{m+1}}\!d\xi\mathcal{N}} e^{\!- i \Delta t_{m} \mathcal{L}/2} q_{n}(t_{m}) \label{eqn:FSS_scheme}
\end{align}
where $\mathcal{L} q_{n} = (-\Delta_{n})^{1+\epsilon} q_{n}$ and $\mathcal{N} q_{n} = \pm |q_{n}|^2(q_{n+1} + q_{n-1})$. The particular operator splitting in equation (\ref{eqn:FSS_scheme}) makes the solution method $\mathcal{O}(\Delta t^2)$ accurate \cite{SUZUKI1992387,YOSHIDA1990262}. The linear step, $e^{- i \Delta t_{m} \mathcal{L}/2}$, is evaluated using discrete Fourier transforms, while the nonlinear step, $e^{i \Delta t_{m} \mathcal{N}}$, is evaluated by solving the associated differential equation, equation (\ref{eqn:fADNLS}) with $(-\Delta_{n})^{1+\epsilon} q_{n} \to 0$, using a fourth-order Runge-Kutta scheme. Throughout this manuscript, solutions to the fADNLS equation were computed with the parameters $h = 1$, $\xi = 0.5$, and $\psi = \pi/2$ and with $N = 2,000$ grid points and time discretization $\Delta t = 0.01$.

\begin{figure}
\begin{centering}
\includegraphics[width=0.45\textwidth]{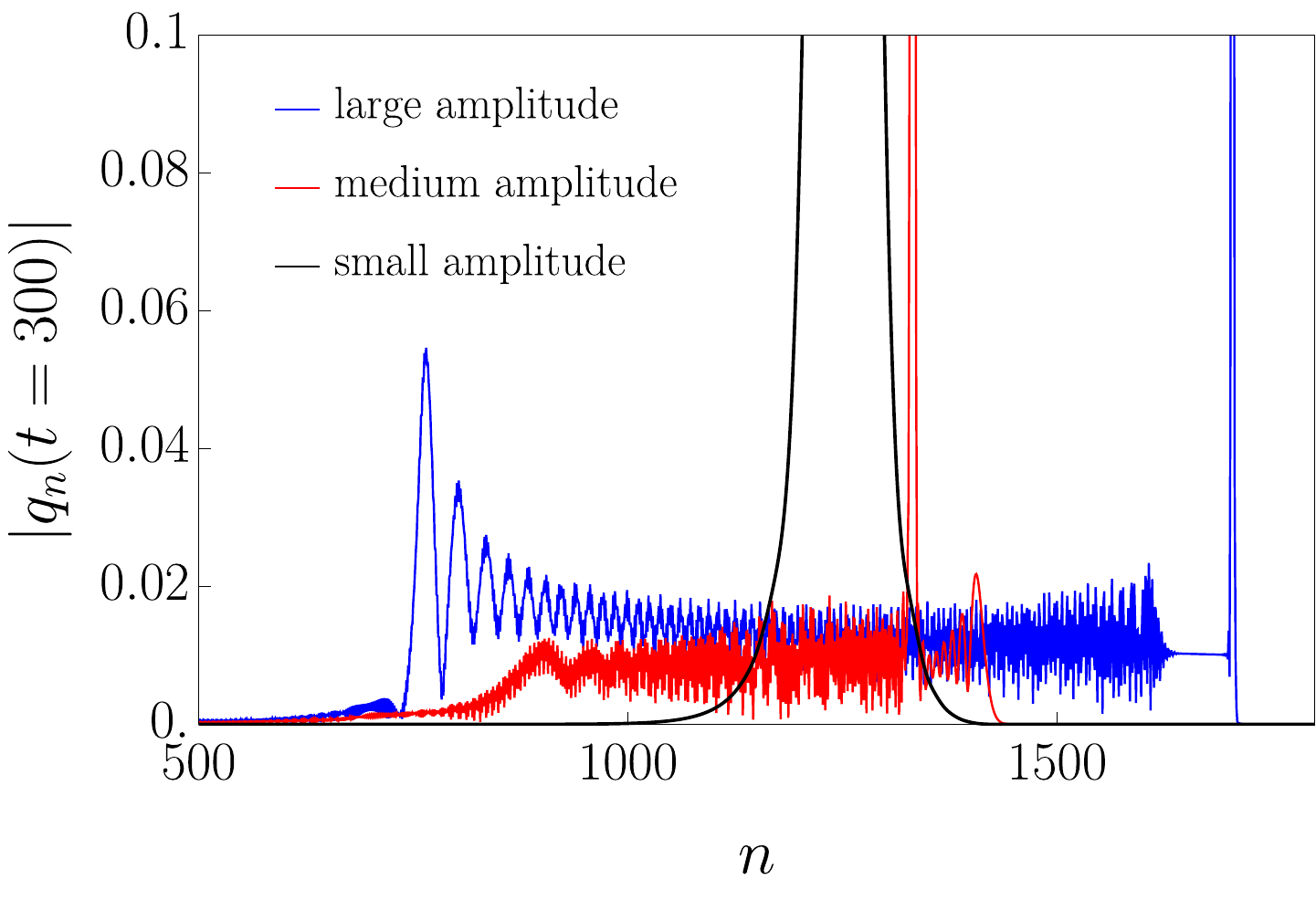}
\caption{\label{fig:radiation_shelf}
    {\it Radiation emission  for small, medium, and large initial data.} Solitary wave solutions to the fractional averaged equation emits more radiation for larger amplitude initial conditions and larger fractional order $\epsilon$ ($\epsilon = 0.1$ is shown here). The initial amplitudes corresponding to the small, medium, and large solutions are $A = 0.100$, $1.175$, and $3.627$, respectively; however, each solitary wave solution is normalized to peak height $1$.}
\end{centering}
\end{figure}

The fADNLS equation initialized with the soliton solution to fIDNLS, i.e., putting $t = 0$ into Eq. (\ref{eqn:fIDNLS_one_soliton}), leads to radiation emission for non-zero $\epsilon$. Figure \ref{fig:radiation_shelf} shows this radiation for small ($\eta = 0.05$), medium ($\eta = 0.5$), and large ($\eta = 1$) amplitude initial conditions at simulation time $T = 300$ with $\epsilon = 0.1$. Recall that amplitude is related to the paramters $\eta$ and $h$ ($h$ is taken to be $1$) by $A = \sinh{(2\eta h)}/h$. The heights of the three solutions are normalized to $1$ to compare the relative amount of radiation; the radiation increases with increasing amplitude, with the large amplitude solution having radiation about $2\%$ of the height of the solution, the medium amplitude having $1\%$, and the small amplitude having negligible radiation.

\begin{figure}
\begin{centering}
\includegraphics[width=0.45\textwidth]{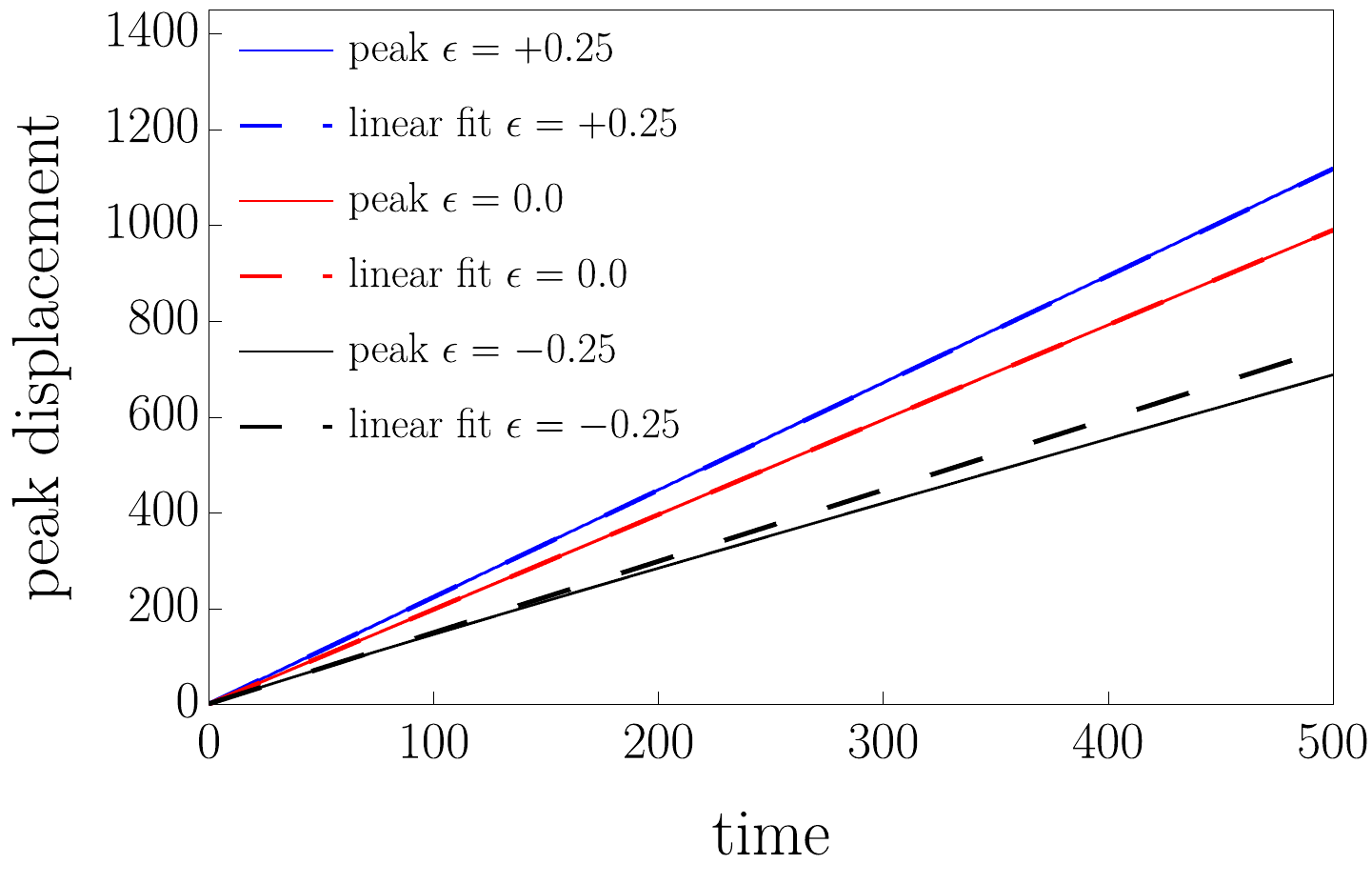}
\caption{\label{fig:peak_position}
    {\it Linearity of solitary wave peak displacement.} Medium amplitude solitary wave solutions of the fractional averaged equation have a nearly linear relationship between displacement and time for positive and zero $\epsilon$.  For negative $\epsilon$, the solitary wave slows down over time.}
\end{centering}
\end{figure}

\begin{figure}
\begin{centering}
\includegraphics[width=0.45\textwidth]{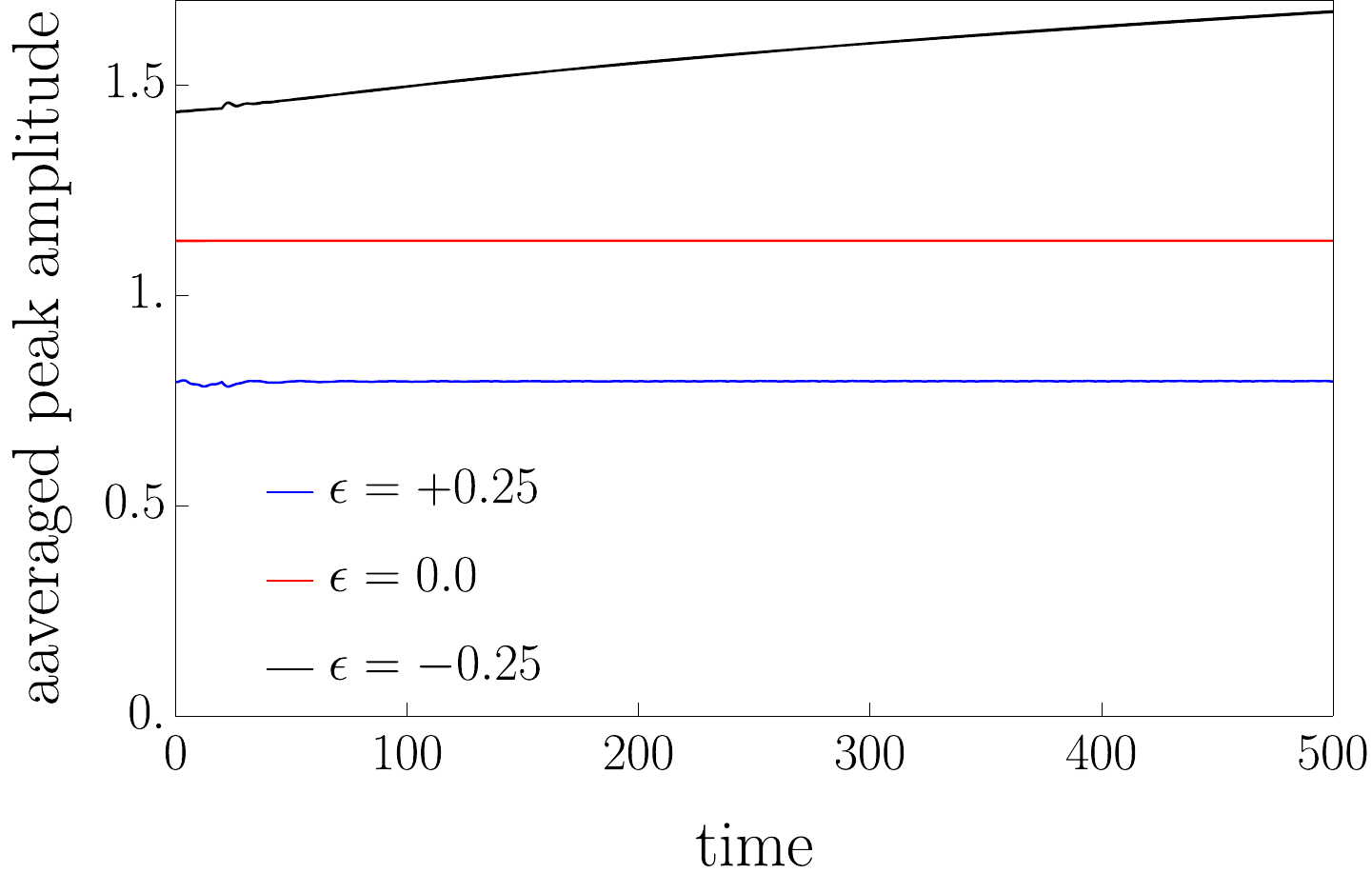}
\caption{\label{fig:peak_amplitude}
    {\it Time-averaged solitary wave peak amplitude.}
    The time-averaged amplitude of medium solitary wave solutions of the fractional averaged equation are nearly constant for positive and zero $\epsilon$ and grow slightly for negative $\epsilon$. The results in this plot and figure \ref{fig:peak_position} suggest that the solitary waves for positive $\epsilon$ are closer to solitons than for negative $\epsilon$.}
\end{centering}
\end{figure}

The positions of the peaks of the fADNLS equation (solid lines) are given along with linear fits (dashed lines) in figure \ref{fig:peak_position} for medium amplitude initial conditions and $\epsilon = -0.25$, $0.0$, $+0.25$. The linear fit shows that the positive $\epsilon$ solution propagates with nearly constant velocity, while the negative $\epsilon$ one has quadratic character which causes it to slow down over time. The amplitudes of these localized wave solutions have breathing patterns. Figure \ref{fig:peak_amplitude} shows that when we average over these oscillations, the amplitude settles down to a constant for $\epsilon = 0.25$ after deformation from the secant profile, but grows a little bit over time for $\epsilon = -0.25$. The averaged amplitude was obtained by taking the mean of the amplitude for $\pm 10$ time units around each point. These results suggest that for $\epsilon$ positive and sufficiently small the localized wave solutions to the fADNLS equation have structure similar to integrable solitons, while those for $\epsilon$ negative are less similar.

\begin{figure}
\begin{centering}
\includegraphics[width=0.45\textwidth]{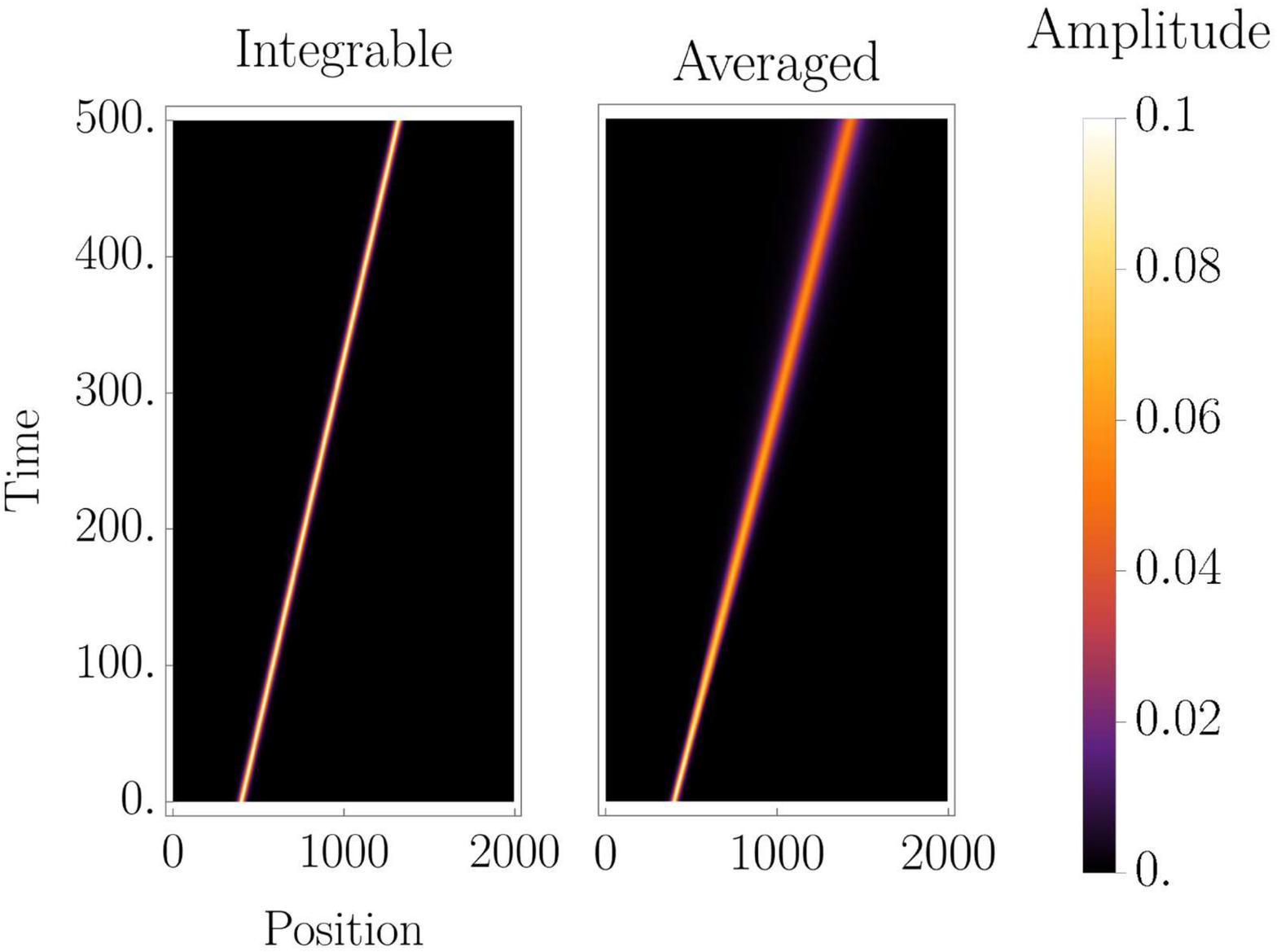}
\caption{\label{fig:density_plot_small}
    {\it Integrable and averaged dynamics for small initial conditions.}
    The soliton solution to the fractional integrable equation propagates at a constant velocity without dissipating for $\epsilon = 0.1$. Even though the profile of the solitary wave solution to the fractional averaged equation deforms from the initial solitonic profile, its peak propagates at a nearly identical velocity to the soliton; the soliton has velocity $1.83864$ and the solitary wave $1.83871 \pm 1 \times 10^{-6}$.}
\end{centering}
\end{figure}

A comparison of a small amplitude soliton solution to the fIDNLS equation and solitary wave solution to the fADNLS equation is given in figure \ref{fig:density_plot_small} for $\epsilon = 0.1$. The solitary wave spreads out, deforming from the hyperbolic secant profile of the soliton. However, the peak velocities of the two waves are nearly identical, $1.83864$ for the soliton and $1.838713 \pm 1 \times 10^{-6}$ for the solitary wave. The soliton moves with exactly constant velocity, but the solitary wave does have an acceleration of $(-1.513 \pm 0.002) \times 10^{-6}$. However, this acceleration is small enough that we can still compare the velocities of these two waves. The velocity and acceleration were estimated by fitting a quadratic curve to the solitary wave peak position and error bounds were obtained by doubling the time discretization, i.e., computing the difference between the results for $\Delta t = 0.01$ and $\Delta t = 0.02$. For larger values of $\epsilon$ and for larger amplitude waves the agreement between these two equations diverges.

\begin{figure}
\begin{centering}
\includegraphics[width=0.45\textwidth]{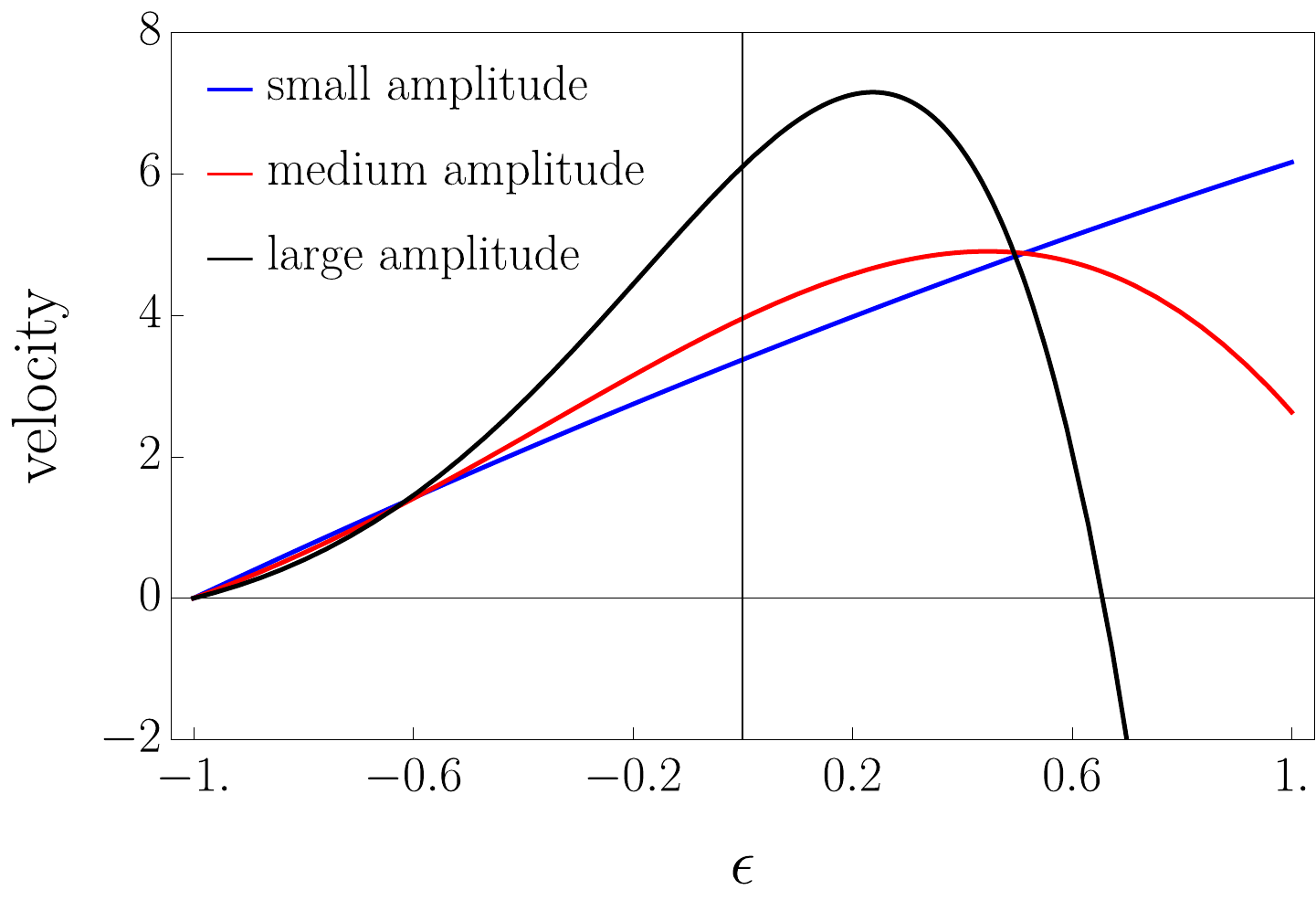}
\caption{\label{fig:integrable_velocity_plot}
    {\it Fractional soliton velocity.}
    Velocity of the one-soliton solution to the fractional integrable equation exhibits super-dispersive transport for small amplitudes ($A = 0.100$). However, for medium ($A = 1.175$) and large ($A = 3.627$) amplitudes, the velocity has a turning point where increasing $\epsilon$ decreases the velocity. This is a fundamentally discrete phenomenon not shared by known continuous fractional integrable equations; cf. \cite{fKdV_fNLS,fmKdV}.}
\end{centering}
\end{figure}

The peak velocity for the one soliton solution to the fIDNLS equation is given by
\begin{align}
    c_{p}(\eta,\xi,h) = \frac{v_{r}}{\eta h}, ~ v_{r} = -2 \text{Im} \big(\sinh^{1+\epsilon}(h[\eta - i \xi]/2)\big) 
\end{align}
which is determined analytically from the form of the soliton in equation (\ref{eqn:fIDNLS_one_soliton}). The peak velocity of the fIDNLS soliton is related to its amplitude in a much more complicated manner than for the fKdV and fNLS equations which have power law relationships between their amplitude and velocity, i.e., anomalous dispersion. Figure \ref{fig:integrable_velocity_plot} shows this velocity as a function of $\epsilon$ for $h = 1$; $\xi = 0.5$; and small, medium, and large amplitudes. 


\section{Conclusion}

In this paper, the fractional integrable discrete nonlinear Schr\"odinger equation was obtained and it's properties were investigated. We did this by applying three principal mathematical constituents which were introduced in our earlier work \cite{fKdV_fNLS}, \cite{fmKdV}: the inverse scattering transform, power law dispersion relations, and completeness relations, to the Ablowitz-Ladik scattering problem. We linearized the equation via Gel'fand-Levitan-Marchenko type summation equations. After long time the kernel of the summation equation contains only discrete spectra; we then obtained an explicit one-soliton solution to this equation, showing that it's velocity depends on the fractional parameter $\epsilon$ in a more complicated way than its continuous counterpart in the fractional nonlinear Schr\"odinger equation. Multi-soliton solutions can be obtained by standard methods; but they are outside the scope of this paper. Using a Fourier split step method, we compared the predictions of the integrable discretization to the fractional averaged nonlinear Schr\"odinger equation, a related non-integrable equation. We demonstrated that for small amplitude initial data, the two equations predicted nearly identical velocities and similar structure, while for large amplitudes they exhibited qualitatively similar characteristics. This work shows that fractional integrability can be substantially extended beyond the continuous nonlinear systems first studied in \cite{fKdV_fNLS}. It suggests new areas of research such as fractional integrability for fully discrete systems. It also opens new opportunities for detailed comparison between fractional nonlinear equations which are integrable to those that are (likely) non-integrable.

\section{Declaration of Competing Interest}

The authors declare that they have no known competing financial interests or personal relationships that could have appeared to
influence the work reported in this paper.

\section{Acknowledgements}
We thank J. Lewis for useful discussions. This project was partially supported by NSF under grants DMS-2005343 and DMR-2002980.

\newpage

\appendix*

\section{Appendix}

\subsection{Scattering Theory for the Ablowitz-Ladik System}

Here, we define eigenfunctions, scattering data, etc. that are used to define the fractional integrable discrete nonlinear Schr\"odinger (fIDNLS) equation and solve it by the IST. The Ablowitz-Ladik scattering problem
\begin{align}
    \mathbf{v}_{n+1} = \begin{pmatrix} z & q_{n} \\ r_{n} & z^{-1} \end{pmatrix} \mathbf{v}_{n} \label{eqn:AL_scattering_problem_SM}
\end{align}
is associated to the following family of nonlinear evolution equations
\begin{align}
    \sigma_{3} \frac{d\mathbf{u}_{n}}{dt} + \gamma(\Lambda_{+}) \mathbf{u}_{n} = 0, ~~ \mathbf{u}_{n} = \left(q_{n}, -r_{n} \right)^{T} \label{eqn:AL_general_evolution_equation_SM}
\end{align}
where $T$ represents transpose, $\sigma_{3} = \text{diag}(1,-1)$, $h_{n} = 1 - r_{n} q_{n}$, $\sum_{n}^{+} = \sum_{k = n}^{\infty}$, and the operator $\Lambda_{+}$ is defined in the main manuscript. Eigenfunctions of the Ablowitz-Ladik scattering system are solutions to equation (\ref{eqn:AL_scattering_problem_SM}) subject to the boundary conditions
\begin{alignat}{3}
    \boldsymbol{\phi}_{n}(z,t) &\sim \begin{pmatrix} z^{n} \\ 0 \end{pmatrix}, ~ &&\overline{\boldsymbol{\phi}}_{n}(z,t) \sim \begin{pmatrix} 0 \\ z^{-n} \end{pmatrix}, ~ &&n\to-\infty, \label{eqn:AL_asymptotic_phi} \\
    \boldsymbol{\psi}_{n}(z,t) &\sim \begin{pmatrix} 0 \\ z^{-n} \end{pmatrix}, ~ &&\overline{\boldsymbol{\psi}}_{n}(z,t) \sim \begin{pmatrix} z^{n} \\ 0 \end{pmatrix}, ~ &&n\to+\infty. \label{eqn:AL_asymptotic_psi}
\end{alignat}
Because the ``right'' eigenfunctions $\boldsymbol{\psi}_{n}$ and $\overline{\boldsymbol{\psi}}_{n}$ are linearly independent, we can write the ``left'' eigenfunctions as
\begin{align}
    \boldsymbol{\phi}_{n}(z,t) &= a(z,t) \overline{\boldsymbol{\psi}}_{n}(z,t) + b(z,t) \boldsymbol{\psi}_{n}(z,t), \label{eqn:AL_scattering_data_phi} \\
    \overline{\boldsymbol{\phi}}_{n}(z,t) &= \overline{a}(z,t) \boldsymbol{\psi}_{n}(z,t) + \overline{b}(z,t) \overline{\boldsymbol{\psi}}_{n}(z,t). \label{eqn:AL_scattering_data_phi_bar}
\end{align}
These relations define the scattering data $a$, $b$, $\overline{a}$, and $\overline{b}$. We can write the scattering data explicitly in terms of the eigenfunctions as
\begin{alignat}{2}
    a(z,t) &= \nu_{n} W\big(\boldsymbol{\phi}_{n},\boldsymbol{\psi}_{n}\big), ~~ &&\overline{a}(z,t) = \nu_{n} W\big(\overline{\boldsymbol{\psi}}_{n},\overline{\boldsymbol{\phi}}_{n}\big), \label{eqn:AL_wronskian_relations_a} \\
    b(z,t) &= \nu_{n} W\big(\overline{\boldsymbol{\psi}}_{n},\boldsymbol{\phi}_{n}\big), ~~ &&\overline{b}(z,t) = \nu_{n} W\big(\overline{\boldsymbol{\phi}}_{n},\boldsymbol{\psi}_{n}\big). \label{eqn:AL_wronskian_relations_b}
\end{alignat}
with the Wronskian $W(\mathbf{u}_{n},\mathbf{v}_{n}) \equiv u_{n}^{(1)} v_{n}^{(2)} - u_{n}^{(2)} v_{n}^{(1)}$ and $\nu_{n} \equiv \prod_{k = n}^{\infty} h_{k}$, $h_{k} = 1 - r_{k} q_{k}$. The transmission and reflection coefficients, $\tau(z,t)$, $\overline{\tau}(z,t)$ and $\rho(z,t)$, $\overline{\rho}(z,t)$, respectively, are defined by
\begin{alignat}{2}
    \tau(z,t) &= \frac{1}{a(z,t)}, \quad &&\rho(z,t) = \frac{b(z,t)}{a(z,t)}, \label{eqn:AL_tau_rho}  \\
    \overline{\tau}(z,t) &= \frac{1}{\overline{a}(z,t)}, \quad &&\overline{\rho}(z,t) = \frac{\overline{b}(z,t)}{\overline{a}(z,t)}. \label{eqn:AL_tau_rho_bar}
\end{alignat}
Often, the functions $\tau$, $\overline{\tau}$, $\rho$, and $\overline{\rho}$ are equivalently referred to as the scattering data. The eigenfunctions
\begin{align}
    \boldsymbol{\phi}_{n}(z,t) z^{-n}, ~~ \boldsymbol{\psi}_{n}(z,t) z^{n}
\end{align}
are analytic and bounded for $|z|>1$ and continuous for $|z|\geq1$ and
\begin{align}
    \overline{\boldsymbol{\phi}}_{n}(z,t) z^{n}, ~~ \overline{\boldsymbol{\psi}}_{n}(z,t) z^{-n}
\end{align}
are analytic and bounded for $|z|<1$ and continuous for $|z|\leq1$. Hence $a$ and $\overline{a}$ are analytic inside and outside the unit circle, respectively.

The Ablowitz-Ladik scattering system can have discrete eigenvalues, corresponding to bound states. These occur at the zeros of $a$ and $\overline{a}$ --- which we notate by $z_{j}$ for $j = 1, 2, ..., J$ and $\overline{z}_{j}$ for $j = 1, 2, ..., \overline{J}$, respectively --- such that $|z_{j}| > 1$ and $|\overline{z}_{j}| < 1$. We assume that these eigenvalues are proper, i.e., the zeros of $a$ and $\overline{a}$ are simple (not on the unit circle and finite in number). At these discrete eigenvalues, the eigenfunctions are related by
\begin{align}
    \boldsymbol{\phi}_{n}(z_{j},t) = b(z_{j},t) \boldsymbol{\psi}_{n}(z_{j},t), ~~ \overline{\boldsymbol{\phi}}_{n}(\overline{z}_{j},t) = \overline{b}(\overline{z}_{j},t) \overline{\boldsymbol{\psi}}_{n}(\overline{z}_{j},t).
\end{align}
We also define the norming constants by
\begin{align}
    c_{j}(t) = \frac{b(z_{j},t)}{a'(z_{j},t)}, ~~ \overline{c}_{j}(t) = \frac{\overline{b}(\overline{z}_{j},t)}{\overline{a}'(\overline{z}_{j},t)}
\end{align}
where $a'(z_{j},t) = \partial_{z} a(z,t) |_{z = z_{j}}$, etc. When $r_{n} = \mp q_{n}^{*}$ in (\ref{eqn:AL_scattering_problem_SM}), we have the symmetry reductions
\begin{align}
    \overline{\boldsymbol{\phi}}_{n}(z,t) = \mathbf{P}_{\mp} \boldsymbol{\phi}_{n}^{*}(1/z^{*},t), ~~ \overline{\boldsymbol{\psi}}_{n}(z,t) = \mp \mathbf{P}_{\mp} \boldsymbol{\psi}_{n}^{*}(1/z^{*},t)
\end{align}
for the eigenfunctions and $\overline{a}(z,t) = a^{*}(1/z^{*},t)$ and $\overline{b}(z,t) = \mp b^{*}(1/z^{*},t)$ where
\begin{align}
    \mathbf{P}_{\mp} = \begin{pmatrix} 0 & \mp 1 \\ 1 & 0 \end{pmatrix}.
\end{align}
The relation $\overline{a}(z,t) = a^{*}(1/z^{*},t)$ implies that if $z_j$ is a zero (eigenvalue) of $a(z,t)$, then $\overline{z}_j = 1/z_{j}^{*}$, $j = 1, 2, ...$ is a zero of $\overline{a}(z,t)$ and hence $J = \overline{J}$. From the eigenfunctions, solutions of (\ref{eqn:AL_scattering_problem_SM}), we can construct the eigenfunctions of the nonlinear operator $\Lambda_{+}$, $\mathbf{\Psi}_{n}(z,t)$ and $\overline{\mathbf{\Psi}}_{n}(z,t)$, and its adjoint $\Lambda_{+}^{A}$, $\mathbf{\Psi}_{n}^{A}(z,t)$ and $\overline{\mathbf{\Psi}}_{n}^{A}(z,t)$ by
\begin{alignat}{2}
    \mathbf{\Psi}_{n} &= \nu_{n} \boldsymbol{\psi}_{n} \circ \boldsymbol{\psi}_{n+1}, ~~ &&\mathbf{\Psi}^{A}_{n} = - \nu_{n} \mathbf{P}_{-} (\boldsymbol{\phi}_{n} \circ \boldsymbol{\phi}_{n+1}) \\
    \overline{\mathbf{\Psi}}_{n} &= \nu_{n} \overline{\boldsymbol{\psi}}_{n} \circ \overline{\boldsymbol{\psi}}_{n+1}, ~~ &&\overline{\mathbf{\Psi}}^{A}_{n} = - \nu_{n} \mathbf{P}_{-} (\overline{\boldsymbol{\phi}}_{n} \circ \overline{\boldsymbol{\phi}}_{n+1})
\end{alignat}
where $\mathbf{u}_{n} \circ \mathbf{v}_{m} = \left( u_{n}^{(1)} v_{m}^{(1)}, u_{n}^{(2)} v_{m}^{(2)} \right)^{T}$.

\subsection{Solving The Nonlinear Evolution Equation Using the IST}

Solving nonlinear discrete evolution equations with the IST is analogous to solving linear discrete evolution equations with the Z-transform. The IST has three distinct steps: direct scattering, time evolution, and inverse scattering which are analagous to taking the Z-transform, evolving the solution in frequency space, and taking the inverse Z-transform, respectively. In direct scattering, the initial condition is mapped into scattering space by solving the scattering problem (\ref{eqn:AL_scattering_problem_SM}). The time evolution of the scattering data, which represents the solution in scattering space, is evolved in time by solving a simple set of differential equations. Finally, in inverse scattering, the solution in physical space is reconstructed from the scattering data by solving a system of algebraic and summation equations. In the following, we briefly outline direct scattering, time evolution, and inverse scattering for the Ablowitz-Ladik scattering system.

\subsubsection{Direct Scattering}

To perform direct scattering, we use the scattering problem in (\ref{eqn:AL_scattering_problem_SM}) to solve for the eigenfunctions $\boldsymbol{\phi}$, $\overline{\boldsymbol{\phi}}$, $\boldsymbol{\psi}$, and $\overline{\boldsymbol{\psi}}$ at $t = 0$. Existence and uniqueness of these solutions can be proven by converting equation (\ref{eqn:AL_scattering_problem_SM}) and the appropriate boundary conditions into linear summation equations which have uniformly convergent Neumann series \cite{Ablowitz3}. These series also provide an alternative method of constructing these eigenfunctions. Then, the scattering data, $a$, $b$, $\overline{a}$, and $\overline{b}$, at $t = 0$ are obtained from the Wronskian relations in equations (\ref{eqn:AL_wronskian_relations_a}) and (\ref{eqn:AL_wronskian_relations_b}).

\subsubsection{Time Evolution}

The scattering data evolves in time according to \cite{gerdjikov1984expansions}
\begin{alignat}{2}
    &\frac{d \rho}{d t} - \gamma(z^2) \rho(z,t) = 0, ~~ &&\frac{d \overline{\rho}}{d t} + \gamma(z^2) \overline{\rho}(z,t) = 0, \label{eqn:AL_scattering_data_time_evolution_rho} \\
    &\frac{d c_{j}}{d t} - \gamma(z_{j}^2) c_{j}(t) = 0, ~~ &&\frac{d \overline{c}_{j}}{d t} + \gamma(\overline{z}_{j}^2) \overline{c}_{j}(t) = 0 \label{eqn:AL_scattering_data_time_evolution_c}
\end{alignat}
for $j = 1, 2, ..., J$ and $j = 1, 2, ..., \overline{J}$, respectively. We recall that $\gamma$ is the function of an operator in equation (\ref{eqn:AL_general_evolution_equation_SM}) and is related to a linear dispersion relation. Also note that $z_{j}$ and $\overline{z}_{j}$ are independent of time. To fully characterized the spectral representation of the operator $\gamma(\Lambda_{+})$, and find the solution $q_n(t)$, we need the eigenfunctions at time $t$ in addition to the scattering data. These functions are found using inverse scattering.

\subsubsection{Inverse Scattering}

To reconstruct the solutions to the nonlinear evolution equation (\ref{eqn:AL_general_evolution_equation_SM}) and eigenfunctions at time $t$, we solve the following Gel'fand-Levitan-Marchenko (GLM) type summation equations for $\boldsymbol{\kappa}(n,m,t)$ \cite{Ablowitz3}
\begin{align}
    \boldsymbol{\kappa}(n,m,t) &+ \begin{pmatrix} 1 \\ 0 \end{pmatrix} \overline{F}(m+n,t) \label{eqn:inverse_scattering_GLM_integral_equation_kappa} \\
    &+ \sum_{j = n+1}^{\infty} \overline{\boldsymbol{\kappa}}(n,j,t) \overline{F}(m+j,t) = 0, \notag \\
    \overline{\boldsymbol{\kappa}}(n,m,t) &+ \begin{pmatrix} 0 \\ 1 \end{pmatrix} F(m+n,t) \label{eqn:inverse_scattering_GLM_integral_equation_kappa_bar} \\ 
    &+ \sum_{j = n+1}^{\infty} \boldsymbol{\kappa}(n,j,t) F(m+j,t) = 0 \notag
\end{align}
where
\begin{align}
    F(n,t) &= \sum_{j = 1}^{J} z_{j}^{-n-1} c_{j}(t) + \frac{1}{2 \pi i} \oint_{S_{1}}\! z^{-n-1} \rho(z,t) dz, \label{eqn:GLM_F} \\
    \overline{F}(n,t) &= \sum_{j = 1}^{\overline{J}} \overline{z}_{j}^{-n-1} \overline{c}_{j}(t) + \frac{1}{2 \pi i} \oint_{S_{1}} \! z^{n-1} \overline{\rho}(z,t) dz. \label{eqn:GLM_Fbar}
\end{align}
Then, the potentials can be obtained from
\begin{align}
    q_{n}(t) = - \kappa^{(1)}(n,n+1,t), ~~ r_{n}(t) = - \overline{\kappa}^{(2)}(n,n+1,t),
\end{align}
and the right eigenfunctions from
\begin{align}
    \boldsymbol{\psi}_{n}(z,t) &= \sum_{j = n}^{\infty} z^{-j} \mathbf{K}(n,j,t), \\ \overline{\boldsymbol{\psi}}_{n}(z,t) &= \sum_{j = n}^{\infty} z^{j} \overline{\mathbf{K}}(n,j,t)
\end{align}
where
\begin{align}
    \mathbf{K}(n,m,t) &= \nu_{n} \boldsymbol{\kappa}(n,m,t), \\
    \overline{\mathbf{K}}(n,m,t) &= \nu_{n} \overline{\boldsymbol{\kappa}}(n,m,t).
\end{align}
The left eigenfunctions $\boldsymbol{\phi}_{n}(z,t)$ and $\overline{\boldsymbol{\phi}}_{n}(z,t)$ can be constructed using the relations in equations (\ref{eqn:AL_scattering_data_phi}) and (\ref{eqn:AL_scattering_data_phi_bar}). If $r_{n}(t) = \mp q_{n}^{*}(t)$, then equations (\ref{eqn:inverse_scattering_GLM_integral_equation_kappa}) and (\ref{eqn:inverse_scattering_GLM_integral_equation_kappa_bar}) both reduce to
\begin{align}
    \kappa^{(1)}(n,m,t) &- \overline{F}(n+m,t) \pm\!\!\!\sum_{n'' = n+1}^{\infty} \sum_{n' = n+1}^{\infty}\! \kappa^{(1)}(n,n'',t) \notag \\
    &\cdot \overline{F}^{*}(n''+n',t) \overline{F}(n'+m,t) = 0
\end{align}
We note that under $r_{n}(t) = \mp q_{n}^{*}(t)$ there are induced symmetries: $\overline{\rho}(z) = \mp \rho^{*}(1/z^{*})$ and for $r_{n}(t) = - q_{n}^{*}(t)$ there can be discrete states with $\overline{z}_j = 1/z_j^{*}$ (hence $\overline{J} = J$), $\overline{c}_{j} = (z_j^{*})^{-2}c_{j}^{*}$, $j = 1, 2, ..., J$. The above GLM summation equations provide a linearization of the fIDNLS equation. Moreover, as $t\to\infty$, the integral terms in the kernels $F$, $\overline{F}$ given by equations (\ref{eqn:GLM_F}) and (\ref{eqn:GLM_Fbar}) vanish. Hence, we are left with only discrete spectra which yields the multisoliton solutions.

\bibliography{refs}

\begin{thebibliography}{54}%
\makeatletter
\providecommand \@ifxundefined [1]{%
 \@ifx{#1\undefined}
}%
\providecommand \@ifnum [1]{%
 \ifnum #1\expandafter \@firstoftwo
 \else \expandafter \@secondoftwo
 \fi
}%
\providecommand \@ifx [1]{%
 \ifx #1\expandafter \@firstoftwo
 \else \expandafter \@secondoftwo
 \fi
}%
\providecommand \natexlab [1]{#1}%
\providecommand \enquote  [1]{``#1''}%
\providecommand \bibnamefont  [1]{#1}%
\providecommand \bibfnamefont [1]{#1}%
\providecommand \citenamefont [1]{#1}%
\providecommand \href@noop [0]{\@secondoftwo}%
\providecommand \href [0]{\begingroup \@sanitize@url \@href}%
\providecommand \@href[1]{\@@startlink{#1}\@@href}%
\providecommand \@@href[1]{\endgroup#1\@@endlink}%
\providecommand \@sanitize@url [0]{\catcode `\\12\catcode `\$12\catcode
  `\&12\catcode `\#12\catcode `\^12\catcode `\_12\catcode `\%12\relax}%
\providecommand \@@startlink[1]{}%
\providecommand \@@endlink[0]{}%
\providecommand \url  [0]{\begingroup\@sanitize@url \@url }%
\providecommand \@url [1]{\endgroup\@href {#1}{\urlprefix }}%
\providecommand \urlprefix  [0]{URL }%
\providecommand \Eprint [0]{\href }%
\providecommand \doibase [0]{https://doi.org/}%
\providecommand \selectlanguage [0]{\@gobble}%
\providecommand \bibinfo  [0]{\@secondoftwo}%
\providecommand \bibfield  [0]{\@secondoftwo}%
\providecommand \translation [1]{[#1]}%
\providecommand \BibitemOpen [0]{}%
\providecommand \bibitemStop [0]{}%
\providecommand \bibitemNoStop [0]{.\EOS\space}%
\providecommand \EOS [0]{\spacefactor3000\relax}%
\providecommand \BibitemShut  [1]{\csname bibitem#1\endcsname}%
\let\auto@bib@innerbib\@empty
\bibitem [{\citenamefont {Korteweg}\ and\ \citenamefont
  {De~Vries}(1895)}]{KDV}%
  \BibitemOpen
  \bibfield  {author} {\bibinfo {author} {\bibfnamefont {D.~J.}\ \bibnamefont
  {Korteweg}}\ and\ \bibinfo {author} {\bibfnamefont {G.}~\bibnamefont
  {De~Vries}},\ }\bibfield  {title} {\bibinfo {title} {Xli. on the change of
  form of long waves advancing in a rectangular canal, and on a new type of
  long stationary waves},\ }\href@noop {} {\bibfield  {journal} {\bibinfo
  {journal} {The London, Edinburgh, and Dublin Philosophical Magazine and
  Journal of Science}\ }\textbf {\bibinfo {volume} {39}},\ \bibinfo {pages}
  {422} (\bibinfo {year} {1895})}\BibitemShut {NoStop}%
\bibitem [{\citenamefont {Ablowitz}\ and\ \citenamefont
  {Segur}(1981)}]{Ablowitz2}%
  \BibitemOpen
  \bibfield  {author} {\bibinfo {author} {\bibfnamefont {M.~J.}\ \bibnamefont
  {Ablowitz}}\ and\ \bibinfo {author} {\bibfnamefont {H.}~\bibnamefont
  {Segur}},\ }\href@noop {} {\emph {\bibinfo {title} {Solitons and the inverse
  scattering transform}}}\ (\bibinfo  {publisher} {SIAM},\ \bibinfo {year}
  {1981})\BibitemShut {NoStop}%
\bibitem [{\citenamefont {Ablowitz}(2011)}]{Ablowitz2011}%
  \BibitemOpen
  \bibfield  {author} {\bibinfo {author} {\bibfnamefont {M.~J.}\ \bibnamefont
  {Ablowitz}},\ }\href@noop {} {\emph {\bibinfo {title} {Nonlinear dispersive
  waves: asymptotic analysis and solitons}}},\ Vol.~\bibinfo {volume} {47}\
  (\bibinfo  {publisher} {Cambridge University Press},\ \bibinfo {year}
  {2011})\BibitemShut {NoStop}%
\bibitem [{\citenamefont {Bronski}\ \emph {et~al.}(2001)\citenamefont
  {Bronski}, \citenamefont {Carr}, \citenamefont {Deconinck},\ and\
  \citenamefont {Kutz}}]{bronski_2001}%
  \BibitemOpen
  \bibfield  {author} {\bibinfo {author} {\bibfnamefont {J.~C.}\ \bibnamefont
  {Bronski}}, \bibinfo {author} {\bibfnamefont {L.~D.}\ \bibnamefont {Carr}},
  \bibinfo {author} {\bibfnamefont {B.}~\bibnamefont {Deconinck}},\ and\
  \bibinfo {author} {\bibfnamefont {J.~N.}\ \bibnamefont {Kutz}},\ }\bibfield
  {title} {\bibinfo {title} {Bose-einstein condensates in standing waves: The
  cubic nonlinear schr\"odinger equation with a periodic potential},\ }\href
  {https://doi.org/10.1103/PhysRevLett.86.1402} {\bibfield  {journal} {\bibinfo
   {journal} {Phys. Rev. Lett.}\ }\textbf {\bibinfo {volume} {86}},\ \bibinfo
  {pages} {1402} (\bibinfo {year} {2001})}\BibitemShut {NoStop}%
\bibitem [{\citenamefont {Boardman}\ \emph {et~al.}(1994)\citenamefont
  {Boardman}, \citenamefont {Wang}, \citenamefont {Nikitov}, \citenamefont
  {Shen}, \citenamefont {Chen}, \citenamefont {Mills},\ and\ \citenamefont
  {Bao}}]{boardman_1994}%
  \BibitemOpen
  \bibfield  {author} {\bibinfo {author} {\bibfnamefont {A.}~\bibnamefont
  {Boardman}}, \bibinfo {author} {\bibfnamefont {Q.}~\bibnamefont {Wang}},
  \bibinfo {author} {\bibfnamefont {S.}~\bibnamefont {Nikitov}}, \bibinfo
  {author} {\bibfnamefont {J.}~\bibnamefont {Shen}}, \bibinfo {author}
  {\bibfnamefont {W.}~\bibnamefont {Chen}}, \bibinfo {author} {\bibfnamefont
  {D.}~\bibnamefont {Mills}},\ and\ \bibinfo {author} {\bibfnamefont
  {J.}~\bibnamefont {Bao}},\ }\bibfield  {title} {\bibinfo {title} {Nonlinear
  magnetostatic surface waves in ferromagnetic films},\ }\href@noop {}
  {\bibfield  {journal} {\bibinfo  {journal} {IEEE transactions on magnetics}\
  }\textbf {\bibinfo {volume} {30}},\ \bibinfo {pages} {14} (\bibinfo {year}
  {1994})}\BibitemShut {NoStop}%
\bibitem [{\citenamefont {Ablowitz}\ \emph {et~al.}(1991)\citenamefont
  {Ablowitz}, \citenamefont {Clarkson},\ and\ \citenamefont
  {Clarkson}}]{Ablowitz1}%
  \BibitemOpen
  \bibfield  {author} {\bibinfo {author} {\bibfnamefont {M.}~\bibnamefont
  {Ablowitz}}, \bibinfo {author} {\bibfnamefont {P.}~\bibnamefont {Clarkson}},\
  and\ \bibinfo {author} {\bibfnamefont {P.~A.}\ \bibnamefont {Clarkson}},\
  }\href@noop {} {\emph {\bibinfo {title} {Solitons, nonlinear evolution
  equations and inverse scattering}}},\ Vol.\ \bibinfo {volume} {149}\
  (\bibinfo  {publisher} {Cambridge university press},\ \bibinfo {year}
  {1991})\BibitemShut {NoStop}%
\bibitem [{\citenamefont {Ablowitz}\ \emph
  {et~al.}(2022{\natexlab{a}})\citenamefont {Ablowitz}, \citenamefont {Been},\
  and\ \citenamefont {Carr}}]{fKdV_fNLS}%
  \BibitemOpen
  \bibfield  {author} {\bibinfo {author} {\bibfnamefont {M.}~\bibnamefont
  {Ablowitz}}, \bibinfo {author} {\bibfnamefont {J.}~\bibnamefont {Been}},\
  and\ \bibinfo {author} {\bibfnamefont {L.}~\bibnamefont {Carr}},\ }\bibfield
  {title} {\bibinfo {title} {Fractional integrable nonlinear soliton
  equations},\ }\href@noop {} {\bibfield  {journal} {\bibinfo  {journal} {Phys.
  Rev. Lett.}\ }\textbf {\bibinfo {volume} {128}},\ \bibinfo {pages} {184101}
  (\bibinfo {year} {2022}{\natexlab{a}})}\BibitemShut {NoStop}%
\bibitem [{\citenamefont {Ablowitz}\ \emph
  {et~al.}(2022{\natexlab{b}})\citenamefont {Ablowitz}, \citenamefont {Been},\
  and\ \citenamefont {Carr}}]{fmKdV}%
  \BibitemOpen
  \bibfield  {author} {\bibinfo {author} {\bibfnamefont {M.~J.}\ \bibnamefont
  {Ablowitz}}, \bibinfo {author} {\bibfnamefont {J.~B.}\ \bibnamefont {Been}},\
  and\ \bibinfo {author} {\bibfnamefont {L.~D.}\ \bibnamefont {Carr}},\
  }\bibfield  {title} {\bibinfo {title} {Integrable fractional modified
  korteweg{\textendash}{deVries}, sine-gordon, and sinh-gordon equations},\
  }\href {https://doi.org/10.1088/1751-8121/ac8844} {\bibfield  {journal}
  {\bibinfo  {journal} {Journal of Physics A: Mathematical and Theoretical}\
  }\textbf {\bibinfo {volume} {55}},\ \bibinfo {pages} {384010} (\bibinfo
  {year} {2022}{\natexlab{b}})}\BibitemShut {NoStop}%
\bibitem [{\citenamefont {West}(2014)}]{west_colloquium}%
  \BibitemOpen
  \bibfield  {author} {\bibinfo {author} {\bibfnamefont {B.~J.}\ \bibnamefont
  {West}},\ }\bibfield  {title} {\bibinfo {title} {Colloquium: Fractional
  calculus view of complexity: A tutorial},\ }\href
  {https://doi.org/10.1103/RevModPhys.86.1169} {\bibfield  {journal} {\bibinfo
  {journal} {Rev. Mod. Phys.}\ }\textbf {\bibinfo {volume} {86}},\ \bibinfo
  {pages} {1169} (\bibinfo {year} {2014})}\BibitemShut {NoStop}%
\bibitem [{\citenamefont {Zhong}\ \emph {et~al.}(2016)\citenamefont {Zhong},
  \citenamefont {Beli{\'c}}, \citenamefont {Malomed}, \citenamefont {Zhang},\
  and\ \citenamefont {Huang}}]{zhong2016spatiotemporal}%
  \BibitemOpen
  \bibfield  {author} {\bibinfo {author} {\bibfnamefont {W.~P.}\ \bibnamefont
  {Zhong}}, \bibinfo {author} {\bibfnamefont {M.~R.}\ \bibnamefont
  {Beli{\'c}}}, \bibinfo {author} {\bibfnamefont {B.~A.}\ \bibnamefont
  {Malomed}}, \bibinfo {author} {\bibfnamefont {Y.}~\bibnamefont {Zhang}},\
  and\ \bibinfo {author} {\bibfnamefont {T.}~\bibnamefont {Huang}},\ }\bibfield
   {title} {\bibinfo {title} {Spatiotemporal accessible solitons in fractional
  dimensions},\ }\href@noop {} {\bibfield  {journal} {\bibinfo  {journal}
  {Physical Review E}\ }\textbf {\bibinfo {volume} {94}},\ \bibinfo {pages}
  {012216} (\bibinfo {year} {2016})}\BibitemShut {NoStop}%
\bibitem [{\citenamefont {Metzler}\ and\ \citenamefont
  {Klafter}(2000)}]{random_walk}%
  \BibitemOpen
  \bibfield  {author} {\bibinfo {author} {\bibfnamefont {R.}~\bibnamefont
  {Metzler}}\ and\ \bibinfo {author} {\bibfnamefont {J.}~\bibnamefont
  {Klafter}},\ }\bibfield  {title} {\bibinfo {title} {The random walk's guide
  to anomalous diffusion: a fractional dynamics approach},\ }\href
  {https://doi.org/https://doi.org/10.1016/S0370-1573(00)00070-3} {\bibfield
  {journal} {\bibinfo  {journal} {Physics Reports}\ }\textbf {\bibinfo {volume}
  {339}},\ \bibinfo {pages} {1} (\bibinfo {year} {2000})}\BibitemShut {NoStop}%
\bibitem [{\citenamefont {Lischke}\ \emph {et~al.}(2020)\citenamefont
  {Lischke}, \citenamefont {Pang}, \citenamefont {Gulian}, \citenamefont
  {Song}, \citenamefont {Glusa}, \citenamefont {Zheng}, \citenamefont {Mao},
  \citenamefont {Cai}, \citenamefont {Meerschaert}, \citenamefont {Ainsworth},\
  and\ \citenamefont {Karniadakis}}]{what_is_frac_lap}%
  \BibitemOpen
  \bibfield  {author} {\bibinfo {author} {\bibfnamefont {A.}~\bibnamefont
  {Lischke}}, \bibinfo {author} {\bibfnamefont {G.}~\bibnamefont {Pang}},
  \bibinfo {author} {\bibfnamefont {M.}~\bibnamefont {Gulian}}, \bibinfo
  {author} {\bibfnamefont {F.}~\bibnamefont {Song}}, \bibinfo {author}
  {\bibfnamefont {C.}~\bibnamefont {Glusa}}, \bibinfo {author} {\bibfnamefont
  {X.}~\bibnamefont {Zheng}}, \bibinfo {author} {\bibfnamefont
  {Z.}~\bibnamefont {Mao}}, \bibinfo {author} {\bibfnamefont {W.}~\bibnamefont
  {Cai}}, \bibinfo {author} {\bibfnamefont {M.~M.}\ \bibnamefont
  {Meerschaert}}, \bibinfo {author} {\bibfnamefont {M.}~\bibnamefont
  {Ainsworth}},\ and\ \bibinfo {author} {\bibfnamefont {G.~E.}\ \bibnamefont
  {Karniadakis}},\ }\bibfield  {title} {\bibinfo {title} {What is the
  fractional laplacian? a comparative review with new results},\ }\href
  {https://doi.org/https://doi.org/10.1016/j.jcp.2019.109009} {\bibfield
  {journal} {\bibinfo  {journal} {Journal of Computational Physics}\ }\textbf
  {\bibinfo {volume} {404}},\ \bibinfo {pages} {109009} (\bibinfo {year}
  {2020})}\BibitemShut {NoStop}%
\bibitem [{\citenamefont {Meerschaert}\ and\ \citenamefont
  {Sikorskii}(2011)}]{frac_stoch}%
  \BibitemOpen
  \bibfield  {author} {\bibinfo {author} {\bibfnamefont {M.~M.}\ \bibnamefont
  {Meerschaert}}\ and\ \bibinfo {author} {\bibfnamefont {A.}~\bibnamefont
  {Sikorskii}},\ }\href {https://doi.org/doi:10.1515/9783110258165} {\emph
  {\bibinfo {title} {Stochastic Models for Fractional Calculus}}}\ (\bibinfo
  {publisher} {De Gruyter},\ \bibinfo {year} {2011})\BibitemShut {NoStop}%
\bibitem [{\citenamefont {Shlesinger}\ \emph {et~al.}(1987)\citenamefont
  {Shlesinger}, \citenamefont {West},\ and\ \citenamefont
  {Klafter}}]{west_1987}%
  \BibitemOpen
  \bibfield  {author} {\bibinfo {author} {\bibfnamefont {M.~F.}\ \bibnamefont
  {Shlesinger}}, \bibinfo {author} {\bibfnamefont {B.~J.}\ \bibnamefont
  {West}},\ and\ \bibinfo {author} {\bibfnamefont {J.}~\bibnamefont
  {Klafter}},\ }\bibfield  {title} {\bibinfo {title} {L\'evy dynamics of
  enhanced diffusion: Application to turbulence},\ }\href
  {https://doi.org/10.1103/PhysRevLett.58.1100} {\bibfield  {journal} {\bibinfo
   {journal} {Phys. Rev. Lett.}\ }\textbf {\bibinfo {volume} {58}},\ \bibinfo
  {pages} {1100} (\bibinfo {year} {1987})}\BibitemShut {NoStop}%
\bibitem [{\citenamefont {West}\ \emph {et~al.}(1997)\citenamefont {West},
  \citenamefont {Grigolini}, \citenamefont {Metzler},\ and\ \citenamefont
  {Nonnenmacher}}]{west_1997}%
  \BibitemOpen
  \bibfield  {author} {\bibinfo {author} {\bibfnamefont {B.~J.}\ \bibnamefont
  {West}}, \bibinfo {author} {\bibfnamefont {P.}~\bibnamefont {Grigolini}},
  \bibinfo {author} {\bibfnamefont {R.}~\bibnamefont {Metzler}},\ and\ \bibinfo
  {author} {\bibfnamefont {T.~F.}\ \bibnamefont {Nonnenmacher}},\ }\bibfield
  {title} {\bibinfo {title} {Fractional diffusion and l\'evy stable
  processes},\ }\href {https://doi.org/10.1103/PhysRevE.55.99} {\bibfield
  {journal} {\bibinfo  {journal} {Phys. Rev. E}\ }\textbf {\bibinfo {volume}
  {55}},\ \bibinfo {pages} {99} (\bibinfo {year} {1997})}\BibitemShut {NoStop}%
\bibitem [{\citenamefont {Wang}\ \emph {et~al.}(2020)\citenamefont {Wang},
  \citenamefont {Cherstvy}, \citenamefont {Chechkin}, \citenamefont {Thapa},
  \citenamefont {Seno}, \citenamefont {Liu},\ and\ \citenamefont
  {Metzler}}]{anomalous_issue}%
  \BibitemOpen
  \bibfield  {author} {\bibinfo {author} {\bibfnamefont {W.}~\bibnamefont
  {Wang}}, \bibinfo {author} {\bibfnamefont {A.~G.}\ \bibnamefont {Cherstvy}},
  \bibinfo {author} {\bibfnamefont {A.~V.}\ \bibnamefont {Chechkin}}, \bibinfo
  {author} {\bibfnamefont {S.}~\bibnamefont {Thapa}}, \bibinfo {author}
  {\bibfnamefont {F.}~\bibnamefont {Seno}}, \bibinfo {author} {\bibfnamefont
  {X.}~\bibnamefont {Liu}},\ and\ \bibinfo {author} {\bibfnamefont
  {R.}~\bibnamefont {Metzler}},\ }\bibfield  {title} {\bibinfo {title}
  {Fractional brownian motion with random diffusivity: emerging residual
  nonergodicity below the correlation time},\ }\href
  {https://doi.org/10.1088/1751-8121/aba467} {\bibfield  {journal} {\bibinfo
  {journal} {Journal of Physics A: Mathematical and Theoretical}\ }\textbf
  {\bibinfo {volume} {53}},\ \bibinfo {pages} {474001} (\bibinfo {year}
  {2020})}\BibitemShut {NoStop}%
\bibitem [{\citenamefont {Saxton}(2007)}]{saxton_2007}%
  \BibitemOpen
  \bibfield  {author} {\bibinfo {author} {\bibfnamefont {M.~J.}\ \bibnamefont
  {Saxton}},\ }\bibfield  {title} {\bibinfo {title} {A biological
  interpretation of transient anomalous subdiffusion. i. qualitative model},\
  }\href {https://doi.org/https://doi.org/10.1529/biophysj.106.092619}
  {\bibfield  {journal} {\bibinfo  {journal} {Biophysical Journal}\ }\textbf
  {\bibinfo {volume} {92}},\ \bibinfo {pages} {1178} (\bibinfo {year}
  {2007})}\BibitemShut {NoStop}%
\bibitem [{\citenamefont {Bronstein}\ \emph {et~al.}(2009)\citenamefont
  {Bronstein}, \citenamefont {Israel}, \citenamefont {Kepten}, \citenamefont
  {Mai}, \citenamefont {Shav-Tal}, \citenamefont {Barkai},\ and\ \citenamefont
  {Garini}}]{bronstein_2009}%
  \BibitemOpen
  \bibfield  {author} {\bibinfo {author} {\bibfnamefont {I.}~\bibnamefont
  {Bronstein}}, \bibinfo {author} {\bibfnamefont {Y.}~\bibnamefont {Israel}},
  \bibinfo {author} {\bibfnamefont {E.}~\bibnamefont {Kepten}}, \bibinfo
  {author} {\bibfnamefont {S.}~\bibnamefont {Mai}}, \bibinfo {author}
  {\bibfnamefont {Y.}~\bibnamefont {Shav-Tal}}, \bibinfo {author}
  {\bibfnamefont {E.}~\bibnamefont {Barkai}},\ and\ \bibinfo {author}
  {\bibfnamefont {Y.}~\bibnamefont {Garini}},\ }\bibfield  {title} {\bibinfo
  {title} {Transient anomalous diffusion of telomeres in the nucleus of
  mammalian cells},\ }\href {https://doi.org/10.1103/PhysRevLett.103.018102}
  {\bibfield  {journal} {\bibinfo  {journal} {Phys. Rev. Lett.}\ }\textbf
  {\bibinfo {volume} {103}},\ \bibinfo {pages} {018102} (\bibinfo {year}
  {2009})}\BibitemShut {NoStop}%
\bibitem [{\citenamefont {Weigel}\ \emph {et~al.}(2011)\citenamefont {Weigel},
  \citenamefont {Simon}, \citenamefont {Tamkun},\ and\ \citenamefont
  {Krapf}}]{weigel_2011}%
  \BibitemOpen
  \bibfield  {author} {\bibinfo {author} {\bibfnamefont {A.~V.}\ \bibnamefont
  {Weigel}}, \bibinfo {author} {\bibfnamefont {B.}~\bibnamefont {Simon}},
  \bibinfo {author} {\bibfnamefont {M.~M.}\ \bibnamefont {Tamkun}},\ and\
  \bibinfo {author} {\bibfnamefont {D.}~\bibnamefont {Krapf}},\ }\bibfield
  {title} {\bibinfo {title} {Ergodic and nonergodic processes coexist in the
  plasma membrane as observed by single-molecule tracking},\ }\href@noop {}
  {\bibfield  {journal} {\bibinfo  {journal} {Proceedings of the National
  Academy of Sciences}\ }\textbf {\bibinfo {volume} {108}},\ \bibinfo {pages}
  {6438} (\bibinfo {year} {2011})}\BibitemShut {NoStop}%
\bibitem [{\citenamefont {Regner}\ \emph {et~al.}(2013)\citenamefont {Regner},
  \citenamefont {Vu{\v{c}}ini{\'c}}, \citenamefont {Domnisoru}, \citenamefont
  {Bartol}, \citenamefont {Hetzer}, \citenamefont {Tartakovsky},\ and\
  \citenamefont {Sejnowski}}]{regner_2013}%
  \BibitemOpen
  \bibfield  {author} {\bibinfo {author} {\bibfnamefont {B.~M.}\ \bibnamefont
  {Regner}}, \bibinfo {author} {\bibfnamefont {D.}~\bibnamefont
  {Vu{\v{c}}ini{\'c}}}, \bibinfo {author} {\bibfnamefont {C.}~\bibnamefont
  {Domnisoru}}, \bibinfo {author} {\bibfnamefont {T.~M.}\ \bibnamefont
  {Bartol}}, \bibinfo {author} {\bibfnamefont {M.~W.}\ \bibnamefont {Hetzer}},
  \bibinfo {author} {\bibfnamefont {D.~M.}\ \bibnamefont {Tartakovsky}},\ and\
  \bibinfo {author} {\bibfnamefont {T.~J.}\ \bibnamefont {Sejnowski}},\
  }\bibfield  {title} {\bibinfo {title} {Anomalous diffusion of single
  particles in cytoplasm},\ }\href@noop {} {\bibfield  {journal} {\bibinfo
  {journal} {Biophysical journal}\ }\textbf {\bibinfo {volume} {104}},\
  \bibinfo {pages} {1652} (\bibinfo {year} {2013})}\BibitemShut {NoStop}%
\bibitem [{\citenamefont {Scher}\ and\ \citenamefont
  {Montroll}(1975)}]{scher_1975}%
  \BibitemOpen
  \bibfield  {author} {\bibinfo {author} {\bibfnamefont {H.}~\bibnamefont
  {Scher}}\ and\ \bibinfo {author} {\bibfnamefont {E.~W.}\ \bibnamefont
  {Montroll}},\ }\bibfield  {title} {\bibinfo {title} {Anomalous transit-time
  dispersion in amorphous solids},\ }\href
  {https://doi.org/10.1103/PhysRevB.12.2455} {\bibfield  {journal} {\bibinfo
  {journal} {Phys. Rev. B}\ }\textbf {\bibinfo {volume} {12}},\ \bibinfo
  {pages} {2455} (\bibinfo {year} {1975})}\BibitemShut {NoStop}%
\bibitem [{\citenamefont {Pfister}\ and\ \citenamefont
  {Scher}(1977)}]{pfister_1977}%
  \BibitemOpen
  \bibfield  {author} {\bibinfo {author} {\bibfnamefont {G.}~\bibnamefont
  {Pfister}}\ and\ \bibinfo {author} {\bibfnamefont {H.}~\bibnamefont
  {Scher}},\ }\bibfield  {title} {\bibinfo {title} {Time-dependent electrical
  transport in amorphous solids: ${\mathrm{as}}_{2}$ ${\mathrm{se}}_{3}$},\
  }\href {https://doi.org/10.1103/PhysRevB.15.2062} {\bibfield  {journal}
  {\bibinfo  {journal} {Phys. Rev. B}\ }\textbf {\bibinfo {volume} {15}},\
  \bibinfo {pages} {2062} (\bibinfo {year} {1977})}\BibitemShut {NoStop}%
\bibitem [{\citenamefont {Gu}\ \emph {et~al.}(1996)\citenamefont {Gu},
  \citenamefont {Schiff}, \citenamefont {Grebner}, \citenamefont {Wang},\ and\
  \citenamefont {Schwarz}}]{gu_1996}%
  \BibitemOpen
  \bibfield  {author} {\bibinfo {author} {\bibfnamefont {Q.}~\bibnamefont
  {Gu}}, \bibinfo {author} {\bibfnamefont {E.~A.}\ \bibnamefont {Schiff}},
  \bibinfo {author} {\bibfnamefont {S.}~\bibnamefont {Grebner}}, \bibinfo
  {author} {\bibfnamefont {F.}~\bibnamefont {Wang}},\ and\ \bibinfo {author}
  {\bibfnamefont {R.}~\bibnamefont {Schwarz}},\ }\bibfield  {title} {\bibinfo
  {title} {Non-gaussian transport measurements and the einstein relation in
  amorphous silicon},\ }\href {https://doi.org/10.1103/PhysRevLett.76.3196}
  {\bibfield  {journal} {\bibinfo  {journal} {Phys. Rev. Lett.}\ }\textbf
  {\bibinfo {volume} {76}},\ \bibinfo {pages} {3196} (\bibinfo {year}
  {1996})}\BibitemShut {NoStop}%
\bibitem [{\citenamefont {Benson}\ \emph {et~al.}(2000)\citenamefont {Benson},
  \citenamefont {Wheatcraft},\ and\ \citenamefont {Meerschaert}}]{benson_2000}%
  \BibitemOpen
  \bibfield  {author} {\bibinfo {author} {\bibfnamefont {D.~A.}\ \bibnamefont
  {Benson}}, \bibinfo {author} {\bibfnamefont {S.~W.}\ \bibnamefont
  {Wheatcraft}},\ and\ \bibinfo {author} {\bibfnamefont {M.~M.}\ \bibnamefont
  {Meerschaert}},\ }\bibfield  {title} {\bibinfo {title} {Application of a
  fractional advection-dispersion equation},\ }\href@noop {} {\bibfield
  {journal} {\bibinfo  {journal} {Water resources research}\ }\textbf {\bibinfo
  {volume} {36}},\ \bibinfo {pages} {1403} (\bibinfo {year}
  {2000})}\BibitemShut {NoStop}%
\bibitem [{\citenamefont {Benson}\ \emph {et~al.}(2001)\citenamefont {Benson},
  \citenamefont {Schumer}, \citenamefont {Meerschaert},\ and\ \citenamefont
  {Wheatcraft}}]{benson_2001}%
  \BibitemOpen
  \bibfield  {author} {\bibinfo {author} {\bibfnamefont {D.~A.}\ \bibnamefont
  {Benson}}, \bibinfo {author} {\bibfnamefont {R.}~\bibnamefont {Schumer}},
  \bibinfo {author} {\bibfnamefont {M.~M.}\ \bibnamefont {Meerschaert}},\ and\
  \bibinfo {author} {\bibfnamefont {S.~W.}\ \bibnamefont {Wheatcraft}},\
  }\bibfield  {title} {\bibinfo {title} {Fractional dispersion, l{\'e}vy
  motion, and the made tracer tests},\ }\href@noop {} {\bibfield  {journal}
  {\bibinfo  {journal} {Transport in porous media}\ }\textbf {\bibinfo {volume}
  {42}},\ \bibinfo {pages} {211} (\bibinfo {year} {2001})}\BibitemShut
  {NoStop}%
\bibitem [{\citenamefont {Meerschaert}\ \emph {et~al.}(2008)\citenamefont
  {Meerschaert}, \citenamefont {Zhang},\ and\ \citenamefont
  {Baeumer}}]{meerschaert_2008}%
  \BibitemOpen
  \bibfield  {author} {\bibinfo {author} {\bibfnamefont {M.~M.}\ \bibnamefont
  {Meerschaert}}, \bibinfo {author} {\bibfnamefont {Y.}~\bibnamefont {Zhang}},\
  and\ \bibinfo {author} {\bibfnamefont {B.}~\bibnamefont {Baeumer}},\
  }\bibfield  {title} {\bibinfo {title} {Tempered anomalous diffusion in
  heterogeneous systems},\ }\href@noop {} {\bibfield  {journal} {\bibinfo
  {journal} {Geophysical Research Letters}\ }\textbf {\bibinfo {volume} {35}}
  (\bibinfo {year} {2008})}\BibitemShut {NoStop}%
\bibitem [{\citenamefont {de~Pablo}\ \emph {et~al.}(2011)\citenamefont
  {de~Pablo}, \citenamefont {Quir{\'o}s}, \citenamefont {Rodr{\'\i}guez},\ and\
  \citenamefont {V{\'a}zquez}}]{frac_porous}%
  \BibitemOpen
  \bibfield  {author} {\bibinfo {author} {\bibfnamefont {A.}~\bibnamefont
  {de~Pablo}}, \bibinfo {author} {\bibfnamefont {F.}~\bibnamefont
  {Quir{\'o}s}}, \bibinfo {author} {\bibfnamefont {A.}~\bibnamefont
  {Rodr{\'\i}guez}},\ and\ \bibinfo {author} {\bibfnamefont {J.~L.}\
  \bibnamefont {V{\'a}zquez}},\ }\bibfield  {title} {\bibinfo {title} {A
  fractional porous medium equation},\ }\href@noop {} {\bibfield  {journal}
  {\bibinfo  {journal} {Advances in Mathematics}\ }\textbf {\bibinfo {volume}
  {226}},\ \bibinfo {pages} {1378} (\bibinfo {year} {2011})}\BibitemShut
  {NoStop}%
\bibitem [{\citenamefont {Koscielny-Bunde}\ \emph {et~al.}(1998)\citenamefont
  {Koscielny-Bunde}, \citenamefont {Bunde}, \citenamefont {Havlin},
  \citenamefont {Roman}, \citenamefont {Goldreich},\ and\ \citenamefont
  {Schellnhuber}}]{koscielny_1998}%
  \BibitemOpen
  \bibfield  {author} {\bibinfo {author} {\bibfnamefont {E.}~\bibnamefont
  {Koscielny-Bunde}}, \bibinfo {author} {\bibfnamefont {A.}~\bibnamefont
  {Bunde}}, \bibinfo {author} {\bibfnamefont {S.}~\bibnamefont {Havlin}},
  \bibinfo {author} {\bibfnamefont {H.~E.}\ \bibnamefont {Roman}}, \bibinfo
  {author} {\bibfnamefont {Y.}~\bibnamefont {Goldreich}},\ and\ \bibinfo
  {author} {\bibfnamefont {H.-J.}\ \bibnamefont {Schellnhuber}},\ }\bibfield
  {title} {\bibinfo {title} {Indication of a universal persistence law
  governing atmospheric variability},\ }\href@noop {} {\bibfield  {journal}
  {\bibinfo  {journal} {Physical Review Letters}\ }\textbf {\bibinfo {volume}
  {81}},\ \bibinfo {pages} {729} (\bibinfo {year} {1998})}\BibitemShut
  {NoStop}%
\bibitem [{\citenamefont {Holm}(2019)}]{power_law_attn}%
  \BibitemOpen
  \bibfield  {author} {\bibinfo {author} {\bibfnamefont {S.}~\bibnamefont
  {Holm}},\ }\href@noop {} {\emph {\bibinfo {title} {Waves with power-law
  attenuation}}}\ (\bibinfo  {publisher} {Springer},\ \bibinfo {year}
  {2019})\BibitemShut {NoStop}%
\bibitem [{\citenamefont {Laskin}(2000)}]{laskin2000fractional}%
  \BibitemOpen
  \bibfield  {author} {\bibinfo {author} {\bibfnamefont {N.}~\bibnamefont
  {Laskin}},\ }\bibfield  {title} {\bibinfo {title} {Fractional quantum
  mechanics and l{\'e}vy path integrals},\ }\href@noop {} {\bibfield  {journal}
  {\bibinfo  {journal} {Physics Letters A}\ }\textbf {\bibinfo {volume}
  {268}},\ \bibinfo {pages} {298} (\bibinfo {year} {2000})}\BibitemShut
  {NoStop}%
\bibitem [{\citenamefont {Laskin}(2018)}]{laskin2018}%
  \BibitemOpen
  \bibfield  {author} {\bibinfo {author} {\bibfnamefont {N.}~\bibnamefont
  {Laskin}},\ }\href {https://doi.org/doi:10.1142/10541} {\emph {\bibinfo
  {title} {Fractional Quantum Mechanics}}}\ (\bibinfo  {publisher} {World
  Scientific: Singapore},\ \bibinfo {year} {2018})\BibitemShut {NoStop}%
\bibitem [{\citenamefont {Gardner}\ \emph {et~al.}(1967)\citenamefont
  {Gardner}, \citenamefont {Greene}, \citenamefont {Kruskal},\ and\
  \citenamefont {Miura}}]{KDV1967}%
  \BibitemOpen
  \bibfield  {author} {\bibinfo {author} {\bibfnamefont {C.~S.}\ \bibnamefont
  {Gardner}}, \bibinfo {author} {\bibfnamefont {J.~M.}\ \bibnamefont {Greene}},
  \bibinfo {author} {\bibfnamefont {M.~D.}\ \bibnamefont {Kruskal}},\ and\
  \bibinfo {author} {\bibfnamefont {R.~M.}\ \bibnamefont {Miura}},\ }\bibfield
  {title} {\bibinfo {title} {Method for solving the korteweg-devries
  equation},\ }\href@noop {} {\bibfield  {journal} {\bibinfo  {journal}
  {Physical review letters}\ }\textbf {\bibinfo {volume} {19}},\ \bibinfo
  {pages} {1095} (\bibinfo {year} {1967})}\BibitemShut {NoStop}%
\bibitem [{\citenamefont {Shabat}\ and\ \citenamefont {Zakharov}(1972)}]{ZS72}%
  \BibitemOpen
  \bibfield  {author} {\bibinfo {author} {\bibfnamefont {A.}~\bibnamefont
  {Shabat}}\ and\ \bibinfo {author} {\bibfnamefont {V.}~\bibnamefont
  {Zakharov}},\ }\bibfield  {title} {\bibinfo {title} {Exact theory of
  two-dimensional self-focusing and one-dimensional self-modulation of waves in
  nonlinear media},\ }\href@noop {} {\bibfield  {journal} {\bibinfo  {journal}
  {Soviet physics JETP}\ }\textbf {\bibinfo {volume} {34}},\ \bibinfo {pages}
  {62} (\bibinfo {year} {1972})}\BibitemShut {NoStop}%
\bibitem [{\citenamefont {Ablowitz}\ \emph {et~al.}(1974)\citenamefont
  {Ablowitz}, \citenamefont {Kaup}, \citenamefont {Newell},\ and\ \citenamefont
  {Segur}}]{AKNS}%
  \BibitemOpen
  \bibfield  {author} {\bibinfo {author} {\bibfnamefont {M.~J.}\ \bibnamefont
  {Ablowitz}}, \bibinfo {author} {\bibfnamefont {D.~J.}\ \bibnamefont {Kaup}},
  \bibinfo {author} {\bibfnamefont {A.~C.}\ \bibnamefont {Newell}},\ and\
  \bibinfo {author} {\bibfnamefont {H.}~\bibnamefont {Segur}},\ }\bibfield
  {title} {\bibinfo {title} {The inverse scattering transform-fourier analysis
  for nonlinear problems},\ }\href@noop {} {\bibfield  {journal} {\bibinfo
  {journal} {Studies in Applied Mathematics}\ }\textbf {\bibinfo {volume}
  {53}},\ \bibinfo {pages} {249} (\bibinfo {year} {1974})}\BibitemShut
  {NoStop}%
\bibitem [{\citenamefont {Ablowitz}\ and\ \citenamefont
  {Ladik}(1975)}]{ablowitz1975nonlinear}%
  \BibitemOpen
  \bibfield  {author} {\bibinfo {author} {\bibfnamefont {M.~J.}\ \bibnamefont
  {Ablowitz}}\ and\ \bibinfo {author} {\bibfnamefont {J.~F.}\ \bibnamefont
  {Ladik}},\ }\bibfield  {title} {\bibinfo {title} {Nonlinear differential-
  difference equations},\ }\href@noop {} {\bibfield  {journal} {\bibinfo
  {journal} {Journal of Mathematical Physics}\ }\textbf {\bibinfo {volume}
  {16}},\ \bibinfo {pages} {598} (\bibinfo {year} {1975})}\BibitemShut
  {NoStop}%
\bibitem [{\citenamefont {Ablowitz}\ and\ \citenamefont
  {Ladik}(1976)}]{ablowitz1976nonlinear}%
  \BibitemOpen
  \bibfield  {author} {\bibinfo {author} {\bibfnamefont {M.}~\bibnamefont
  {Ablowitz}}\ and\ \bibinfo {author} {\bibfnamefont {J.}~\bibnamefont
  {Ladik}},\ }\bibfield  {title} {\bibinfo {title} {Nonlinear
  differential--difference equations and fourier analysis},\ }\href@noop {}
  {\bibfield  {journal} {\bibinfo  {journal} {Journal of Mathematical Physics}\
  }\textbf {\bibinfo {volume} {17}},\ \bibinfo {pages} {1011} (\bibinfo {year}
  {1976})}\BibitemShut {NoStop}%
\bibitem [{\citenamefont {Molina}(2020)}]{molina2020two}%
  \BibitemOpen
  \bibfield  {author} {\bibinfo {author} {\bibfnamefont {M.~I.}\ \bibnamefont
  {Molina}},\ }\bibfield  {title} {\bibinfo {title} {The two-dimensional
  fractional discrete nonlinear schr{\"o}dinger equation},\ }\href@noop {}
  {\bibfield  {journal} {\bibinfo  {journal} {Physics Letters A}\ }\textbf
  {\bibinfo {volume} {384}},\ \bibinfo {pages} {126835} (\bibinfo {year}
  {2020})}\BibitemShut {NoStop}%
\bibitem [{\citenamefont {Ciaurri}\ \emph {et~al.}(2015)\citenamefont
  {Ciaurri}, \citenamefont {Roncal}, \citenamefont {Stinga}, \citenamefont
  {Torrea},\ and\ \citenamefont {Varona}}]{ciaurri2015fractional}%
  \BibitemOpen
  \bibfield  {author} {\bibinfo {author} {\bibfnamefont {{\'O}.}~\bibnamefont
  {Ciaurri}}, \bibinfo {author} {\bibfnamefont {L.}~\bibnamefont {Roncal}},
  \bibinfo {author} {\bibfnamefont {P.~R.}\ \bibnamefont {Stinga}}, \bibinfo
  {author} {\bibfnamefont {J.~L.}\ \bibnamefont {Torrea}},\ and\ \bibinfo
  {author} {\bibfnamefont {J.~L.}\ \bibnamefont {Varona}},\ }\bibfield  {title}
  {\bibinfo {title} {Fractional discrete laplacian versus discretized
  fractional laplacian},\ }\href@noop {} {\bibfield  {journal} {\bibinfo
  {journal} {arXiv preprint arXiv:1507.04986}\ } (\bibinfo {year}
  {2015})}\BibitemShut {NoStop}%
\bibitem [{\citenamefont {Huang}\ and\ \citenamefont
  {Oberman}(2014)}]{huang2014numerical}%
  \BibitemOpen
  \bibfield  {author} {\bibinfo {author} {\bibfnamefont {Y.}~\bibnamefont
  {Huang}}\ and\ \bibinfo {author} {\bibfnamefont {A.}~\bibnamefont
  {Oberman}},\ }\bibfield  {title} {\bibinfo {title} {Numerical methods for the
  fractional laplacian: A finite difference-quadrature approach},\ }\href@noop
  {} {\bibfield  {journal} {\bibinfo  {journal} {SIAM Journal on Numerical
  Analysis}\ }\textbf {\bibinfo {volume} {52}},\ \bibinfo {pages} {3056}
  (\bibinfo {year} {2014})}\BibitemShut {NoStop}%
\bibitem [{\citenamefont {Ciaurri}\ \emph {et~al.}(2017)\citenamefont
  {Ciaurri}, \citenamefont {Alastair~Gillespie}, \citenamefont {Roncal},
  \citenamefont {Torrea},\ and\ \citenamefont {Varona}}]{ciaurri2017harmonic}%
  \BibitemOpen
  \bibfield  {author} {\bibinfo {author} {\bibfnamefont {{\'O}.}~\bibnamefont
  {Ciaurri}}, \bibinfo {author} {\bibfnamefont {T.}~\bibnamefont
  {Alastair~Gillespie}}, \bibinfo {author} {\bibfnamefont {L.}~\bibnamefont
  {Roncal}}, \bibinfo {author} {\bibfnamefont {J.~L.}\ \bibnamefont {Torrea}},\
  and\ \bibinfo {author} {\bibfnamefont {J.~L.}\ \bibnamefont {Varona}},\
  }\bibfield  {title} {\bibinfo {title} {Harmonic analysis associated with a
  discrete laplacian},\ }\href@noop {} {\bibfield  {journal} {\bibinfo
  {journal} {Journal d'Analyse Math{\'e}matique}\ }\textbf {\bibinfo {volume}
  {132}},\ \bibinfo {pages} {109} (\bibinfo {year} {2017})}\BibitemShut
  {NoStop}%
\bibitem [{\citenamefont {Iomin}(2021)}]{iomin2021fractional}%
  \BibitemOpen
  \bibfield  {author} {\bibinfo {author} {\bibfnamefont {A.}~\bibnamefont
  {Iomin}},\ }\bibfield  {title} {\bibinfo {title} {Fractional schr{\"o}dinger
  equation in gravitational optics},\ }\href@noop {} {\bibfield  {journal}
  {\bibinfo  {journal} {Modern Physics Letters A}\ }\textbf {\bibinfo {volume}
  {36}},\ \bibinfo {pages} {2140003} (\bibinfo {year} {2021})}\BibitemShut
  {NoStop}%
\bibitem [{\citenamefont {Malomed}(2021)}]{malomed2021optical}%
  \BibitemOpen
  \bibfield  {author} {\bibinfo {author} {\bibfnamefont {B.~A.}\ \bibnamefont
  {Malomed}},\ }\bibfield  {title} {\bibinfo {title} {Optical solitons and
  vortices in fractional media: A mini-review of recent results},\ }in\
  \href@noop {} {\emph {\bibinfo {booktitle} {Photonics}}},\ Vol.~\bibinfo
  {volume} {8}\ (\bibinfo {organization} {Multidisciplinary Digital Publishing
  Institute},\ \bibinfo {year} {2021})\ p.\ \bibinfo {pages} {353}\BibitemShut
  {NoStop}%
\bibitem [{\citenamefont {Qiu}\ \emph {et~al.}(2020)\citenamefont {Qiu},
  \citenamefont {Malomed}, \citenamefont {Mihalache}, \citenamefont {Zhu},
  \citenamefont {Peng},\ and\ \citenamefont {He}}]{qiu2020stabilization}%
  \BibitemOpen
  \bibfield  {author} {\bibinfo {author} {\bibfnamefont {Y.}~\bibnamefont
  {Qiu}}, \bibinfo {author} {\bibfnamefont {B.~A.}\ \bibnamefont {Malomed}},
  \bibinfo {author} {\bibfnamefont {D.}~\bibnamefont {Mihalache}}, \bibinfo
  {author} {\bibfnamefont {X.}~\bibnamefont {Zhu}}, \bibinfo {author}
  {\bibfnamefont {X.}~\bibnamefont {Peng}},\ and\ \bibinfo {author}
  {\bibfnamefont {Y.}~\bibnamefont {He}},\ }\bibfield  {title} {\bibinfo
  {title} {Stabilization of single-and multi-peak solitons in the fractional
  nonlinear schr{\"o}dinger equation with a trapping potential},\ }\href@noop
  {} {\bibfield  {journal} {\bibinfo  {journal} {Chaos, Solitons \& Fractals}\
  }\textbf {\bibinfo {volume} {140}},\ \bibinfo {pages} {110222} (\bibinfo
  {year} {2020})}\BibitemShut {NoStop}%
\bibitem [{\citenamefont {Li}\ \emph {et~al.}(2021)\citenamefont {Li},
  \citenamefont {Malomed},\ and\ \citenamefont {Mihalache}}]{li2021symmetry}%
  \BibitemOpen
  \bibfield  {author} {\bibinfo {author} {\bibfnamefont {P.}~\bibnamefont
  {Li}}, \bibinfo {author} {\bibfnamefont {B.~A.}\ \bibnamefont {Malomed}},\
  and\ \bibinfo {author} {\bibfnamefont {D.}~\bibnamefont {Mihalache}},\
  }\bibfield  {title} {\bibinfo {title} {Symmetry-breaking bifurcations and
  ghost states in the fractional nonlinear schr\"odinger equation with a
  pt-symmetric potential},\ }\href@noop {} {\bibfield  {journal} {\bibinfo
  {journal} {arXiv preprint arXiv:2106.05446}\ } (\bibinfo {year}
  {2021})}\BibitemShut {NoStop}%
\bibitem [{\citenamefont {Al~Khawaja}\ \emph {et~al.}(2018)\citenamefont
  {Al~Khawaja}, \citenamefont {Al-Refai}, \citenamefont {Shchedrin},\ and\
  \citenamefont {Carr}}]{al2018high}%
  \BibitemOpen
  \bibfield  {author} {\bibinfo {author} {\bibfnamefont {U.}~\bibnamefont
  {Al~Khawaja}}, \bibinfo {author} {\bibfnamefont {M.}~\bibnamefont
  {Al-Refai}}, \bibinfo {author} {\bibfnamefont {G.}~\bibnamefont
  {Shchedrin}},\ and\ \bibinfo {author} {\bibfnamefont {L.~D.}\ \bibnamefont
  {Carr}},\ }\bibfield  {title} {\bibinfo {title} {High-accuracy power series
  solutions with arbitrarily large radius of convergence for the fractional
  nonlinear schr{\"o}dinger-type equations},\ }\href@noop {} {\bibfield
  {journal} {\bibinfo  {journal} {Journal of Physics A: Mathematical and
  Theoretical}\ }\textbf {\bibinfo {volume} {51}},\ \bibinfo {pages} {235201}
  (\bibinfo {year} {2018})}\BibitemShut {NoStop}%
\bibitem [{\citenamefont {Ablowitz}\ \emph {et~al.}(2021)\citenamefont
  {Ablowitz}, \citenamefont {Cole}, \citenamefont {Hu},\ and\ \citenamefont
  {Rosenthal}}]{PhysRevE.103.042214}%
  \BibitemOpen
  \bibfield  {author} {\bibinfo {author} {\bibfnamefont {M.~J.}\ \bibnamefont
  {Ablowitz}}, \bibinfo {author} {\bibfnamefont {J.~T.}\ \bibnamefont {Cole}},
  \bibinfo {author} {\bibfnamefont {P.}~\bibnamefont {Hu}},\ and\ \bibinfo
  {author} {\bibfnamefont {P.}~\bibnamefont {Rosenthal}},\ }\bibfield  {title}
  {\bibinfo {title} {Peierls-nabarro barrier effect in nonlinear floquet
  topological insulators},\ }\href
  {https://doi.org/10.1103/PhysRevE.103.042214} {\bibfield  {journal} {\bibinfo
   {journal} {Phys. Rev. E}\ }\textbf {\bibinfo {volume} {103}},\ \bibinfo
  {pages} {042214} (\bibinfo {year} {2021})}\BibitemShut {NoStop}%
\bibitem [{\citenamefont {Gerdjikov}\ \emph {et~al.}(1984)\citenamefont
  {Gerdjikov}, \citenamefont {Ivanov},\ and\ \citenamefont
  {Kulish}}]{gerdjikov1984expansions}%
  \BibitemOpen
  \bibfield  {author} {\bibinfo {author} {\bibfnamefont {V.}~\bibnamefont
  {Gerdjikov}}, \bibinfo {author} {\bibfnamefont {M.}~\bibnamefont {Ivanov}},\
  and\ \bibinfo {author} {\bibfnamefont {P.}~\bibnamefont {Kulish}},\
  }\bibfield  {title} {\bibinfo {title} {Expansions over the "squared"
  solutions and difference evolution equations},\ }\href@noop {} {\bibfield
  {journal} {\bibinfo  {journal} {Journal of mathematical physics}\ }\textbf
  {\bibinfo {volume} {25}},\ \bibinfo {pages} {25} (\bibinfo {year}
  {1984})}\BibitemShut {NoStop}%
\bibitem [{\citenamefont {Chiu}\ and\ \citenamefont
  {Ladik}(1977)}]{ChiuLadik1977}%
  \BibitemOpen
  \bibfield  {author} {\bibinfo {author} {\bibfnamefont {S.-C.}\ \bibnamefont
  {Chiu}}\ and\ \bibinfo {author} {\bibfnamefont {J.}~\bibnamefont {Ladik}},\
  }\bibfield  {title} {\bibinfo {title} {Generating exactly soluble nonlinear
  discrete evolution equations by a generalized wronskian technique},\
  }\href@noop {} {\bibfield  {journal} {\bibinfo  {journal} {J. Math. Phys.}\
  }\textbf {\bibinfo {volume} {18}},\ \bibinfo {pages} {690} (\bibinfo {year}
  {1977})}\BibitemShut {NoStop}%
\bibitem [{\citenamefont {Ablowitz}\ \emph {et~al.}(2004)\citenamefont
  {Ablowitz}, \citenamefont {Prinari},\ and\ \citenamefont
  {Trubatch}}]{Ablowitz3}%
  \BibitemOpen
  \bibfield  {author} {\bibinfo {author} {\bibfnamefont {M.}~\bibnamefont
  {Ablowitz}}, \bibinfo {author} {\bibfnamefont {B.}~\bibnamefont {Prinari}},\
  and\ \bibinfo {author} {\bibfnamefont {A.}~\bibnamefont {Trubatch}},\
  }\href@noop {} {\emph {\bibinfo {title} {Discrete and continuous nonlinear
  Schr{\"o}dinger systems}}},\ Vol.\ \bibinfo {volume} {302}\ (\bibinfo
  {publisher} {Cambridge University Press},\ \bibinfo {year}
  {2004})\BibitemShut {NoStop}%
\bibitem [{\citenamefont {Taha}\ and\ \citenamefont
  {Ablowitz}(1984)}]{TAHA1984203}%
  \BibitemOpen
  \bibfield  {author} {\bibinfo {author} {\bibfnamefont {T.~R.}\ \bibnamefont
  {Taha}}\ and\ \bibinfo {author} {\bibfnamefont {M.~I.}\ \bibnamefont
  {Ablowitz}},\ }\bibfield  {title} {\bibinfo {title} {Analytical and numerical
  aspects of certain nonlinear evolution equations. ii. numerical, nonlinear
  schr\"odinger equation},\ }\href
  {https://doi.org/https://doi.org/10.1016/0021-9991(84)90003-2} {\bibfield
  {journal} {\bibinfo  {journal} {Journal of Computational Physics}\ }\textbf
  {\bibinfo {volume} {55}},\ \bibinfo {pages} {203} (\bibinfo {year}
  {1984})}\BibitemShut {NoStop}%
\bibitem [{\citenamefont {Hardin}(1973)}]{Hardin1973ApplicationOT}%
  \BibitemOpen
  \bibfield  {author} {\bibinfo {author} {\bibfnamefont {R.~H.}\ \bibnamefont
  {Hardin}},\ }\bibfield  {title} {\bibinfo {title} {Application of the
  split-step fourier method to the numerical solution of nonlinear and variable
  coefficient wave equations},\ }\href@noop {} {\bibfield  {journal} {\bibinfo
  {journal} {Siam Review}\ }\textbf {\bibinfo {volume} {15}},\ \bibinfo {pages}
  {423} (\bibinfo {year} {1973})}\BibitemShut {NoStop}%
\bibitem [{\citenamefont {Sinkin}\ \emph {et~al.}(2003)\citenamefont {Sinkin},
  \citenamefont {Holzl\"{o}hner}, \citenamefont {Zweck},\ and\ \citenamefont
  {Menyuk}}]{Sinkin:03}%
  \BibitemOpen
  \bibfield  {author} {\bibinfo {author} {\bibfnamefont {O.~V.}\ \bibnamefont
  {Sinkin}}, \bibinfo {author} {\bibfnamefont {R.}~\bibnamefont
  {Holzl\"{o}hner}}, \bibinfo {author} {\bibfnamefont {J.}~\bibnamefont
  {Zweck}},\ and\ \bibinfo {author} {\bibfnamefont {C.~R.}\ \bibnamefont
  {Menyuk}},\ }\bibfield  {title} {\bibinfo {title} {Optimization of the
  split-step fourier method in modeling optical-fiber communications systems},\
  }\href {http://opg.optica.org/jlt/abstract.cfm?URI=jlt-21-1-61} {\bibfield
  {journal} {\bibinfo  {journal} {J. Lightwave Technol.}\ }\textbf {\bibinfo
  {volume} {21}},\ \bibinfo {pages} {61} (\bibinfo {year} {2003})}\BibitemShut
  {NoStop}%
\bibitem [{\citenamefont {Suzuki}(1992)}]{SUZUKI1992387}%
  \BibitemOpen
  \bibfield  {author} {\bibinfo {author} {\bibfnamefont {M.}~\bibnamefont
  {Suzuki}},\ }\bibfield  {title} {\bibinfo {title} {General theory of
  higher-order decomposition of exponential operators and symplectic
  integrators},\ }\href
  {https://doi.org/https://doi.org/10.1016/0375-9601(92)90335-J} {\bibfield
  {journal} {\bibinfo  {journal} {Physics Letters A}\ }\textbf {\bibinfo
  {volume} {165}},\ \bibinfo {pages} {387} (\bibinfo {year}
  {1992})}\BibitemShut {NoStop}%
\bibitem [{\citenamefont {Yoshida}(1990)}]{YOSHIDA1990262}%
  \BibitemOpen
  \bibfield  {author} {\bibinfo {author} {\bibfnamefont {H.}~\bibnamefont
  {Yoshida}},\ }\bibfield  {title} {\bibinfo {title} {Construction of higher
  order symplectic integrators},\ }\href
  {https://doi.org/https://doi.org/10.1016/0375-9601(90)90092-3} {\bibfield
  {journal} {\bibinfo  {journal} {Physics Letters A}\ }\textbf {\bibinfo
  {volume} {150}},\ \bibinfo {pages} {262} (\bibinfo {year}
  {1990})}\BibitemShut {NoStop}%
\end{thebibliography}%

\end{document}